\documentclass{article}
\usepackage{graphicx}  
\usepackage{amsmath}   
\usepackage[compress]{cite}
\usepackage{amssymb}   
\usepackage{bm} 
\usepackage{dcolumn}
\usepackage{color}
\usepackage{mathrsfs}
\usepackage{amsfonts}
\usepackage{varioref}
\usepackage{textcomp}

\RequirePackage[colorlinks,citecolor=blue,urlcolor=magenta,linkcolor=blue]{hyperref}
\allowdisplaybreaks
\usepackage{array}
\renewcommand{\arraystretch}{1.2}
\newcounter{mnotecount}[section]

\renewcommand{\themnotecount}{\thesection.\arabic{mnotecount}}

\newcommand{\mnotex}[1]
{\protect{\stepcounter{mnotecount}}$^{\mbox{\footnotesize
$
\bullet$\themnotecount}}$ \marginpar{
\raggedright\small\em
$\!\!\!\!\!\!\,\bullet$\themnotecount: #1} }
\newcommand{\Sumanta}[1]{\mnotex{{\bf sc:} {\color{blue}  #1}}}

\newcommand{\Shauvik}[1]{\mnotex{{\bf SB:} {\color{red} #1}}}

\def\DS{Damour-Solodukhin }
\labelformat{section}{Section #1} 
\labelformat{subsection}{Section #1} 
\labelformat{subsubsection}{Section #1}
\labelformat{subsubsubsection}{Section #1}
\labelformat{equation}{Eq.~(#1)} 
\labelformat{figure}{Fig.~#1} 
\labelformat{subfigure}{Fig.~\thefigure#1} 
\labelformat{table}{Table~#1} 
\labelformat{appendix}{Appendix #1}
\newcommand{\be}{\nopagebreak[3]\begin{equation}}
\newcommand{\ee}{\end{equation}}

\newcommand{\ba}{\nopagebreak[3]\begin{eqnarray}}
\newcommand{\ea}{\end{eqnarray}}

\title{\bf Galactic wormholes: Geometry, stability, and echoes}
\author{Shauvik Biswas\footnote{shauvikbiswas2014@gmail.com}$~^{1}$, Chiranjeeb Singha\footnote{chiranjeeb.singha@saha.ac.in}$~^{2}$ and Sumanta Chakraborty\footnote{tpsc@iacs.res.in}$~^{1}$
\\
$^{1}${\small{School of Physical Sciences, Indian Association for the Cultivation of Science, Kolkata-700032, India}}\\
$^{2}${\small{Theory Division, Saha Institute of Nuclear Physics, Kolkata 700064, India}}}
\begin{document}

\maketitle
\begin{abstract}

In this work, we present the environmental effects on wormholes residing in a galaxy. By this, we propose that these wormholes are mimickers of supermassive black holes residing at the galactic centers. In particular, we consider two wormhole spacetimes classes: the Damour-Solodukhin wormhole and the braneworld wormhole. While there is no classical matter model for the Damour-Solodukhin wormhole, the braneworld wormhole, on the other hand, is supported by a scalar-tensor theory on the four-dimensional brane. Intriguingly, it turns out that the presence of a dark matter halo surrounding these wormholes can tame the violations of energy conditions present in generic wormhole spacetimes. Our results also demonstrate that the galactic Damour-Solodukhin wormhole is more stable than its isolated counterpart under linear scalar perturbation, whereas we obtain the opposite behavior for the braneworld wormhole. The perturbation of these wormholes leads to echoes in the ringdown waveform, which are sensitive to the properties of the dark matter halo. To be precise, the time delay between two echoes is affected by the galactic matter environment, and it appears to be a generic effect present for any exotic compact object living in a galaxy. This allows us to identify the galactic parameters, independently from the gravitational wave measurements, if echoes are observed in future generations of gravitational wave detectors. For completeness, we have also analyzed the impact of the galactic environment on the photon sphere, the innermost stable circular orbits, and the shadow radius. It turns out that the dark matter halo indeed affects these locations, with implications for shadow and accretion physics.    
\end{abstract}  
\section{Introduction}\label{sec-1}

Understanding gravity at its extreme is the key to solving one of the most important problems of modern physics, i.e., constructing a self-consistent quantum theory of gravity \cite{tHooft:1978, ortín_2004, Rovelli:2004tv, Bojowald:2010qpa}. The urge for quantizing gravity arises from the fact that though it is well-known how to quantize three of the four fundamental forces, namely, the strong, weak, and electromagnetic, and provide a unified description, so far gravity has remained elusive\footnote{Here, by quantum gravity, we mean quantizing gravity with all of its non-linear aspects included. In the linear regime, however, quantum gravity exists, which can be obtained from the fact that in the linear regime general relativity reduces to a massless spin-2 classical field theory on the flat background \cite{Fierz:1939ix, Padmanabhan:2004xk, Deser:2009fq} and hence can be quantized using known prescriptions \cite{gupta1952quantization}.}. In the weak field regime, the gravitational interaction seems to be described extremely well by Einstein's theory of general relativity \cite{Berti:2015itd, Will:2014kxa, will2020einstein}, however, the dynamics of gravity in the strong field regime is currently being unwrapped \cite{Berti:2015itd}. Recent gravitational wave detections from the merger of binary black holes by the LIGO-VIRGO collaboration \cite{LIGOScientific:2016aoc, LIGOScientific:2017bnn, LIGOScientific:2018mvr, LIGOScientific:2020ibl, LIGOScientific:2020tif, LIGOScientific:2020aai} have opened the window to probe the nature of gravity in the strong field regime, in particular the physics near the horizon. Such an analysis is of utmost importance, given the fact that it is being hotly debated whether we really have a black hole or some ultra-compact object, possibly involving quantum gravitational effects modifying the near horizon structure \cite{Almheiri:2012rt, Mathur:2005zp, Wang:2019rcf, Agullo:2020hxe, Chakraborty:2022zlq}. This leads to the exciting possibility of having horizonless but extremely compact objects mimicking the gravitational wave signals seen by the LIGO-VIRGO collaboration \cite{Cardoso:2014sna, Cardoso:2016rao, Cardoso:2016oxy, Cardoso:2019rvt, Dey:2020lhq, Maggio:2021ans}. 

Such ultra-compact objects often have radii smaller than the Buchdahl radius \cite{Buchdahl:1959zz, Dadhich:2016fku, Alho:2021sli, Alho:2022bki}, the limiting stellar configuration achieved using `normal' matter fields, and hence to model such compact objects we need exotic matter fields. Following this, such ultra-compact objects are often referred to as exotic compact objects, or, ECOs. To model these ECOs one usually puts a partially reflecting surface slightly away from the location of the would-be horizon, such that the reflective surface is within the photon sphere, so they can mimic the ringdown waveform of a black hole \cite{Cardoso:2014sna, Cardoso:2016rao, Cardoso:2016oxy, Mark:2017dnq, Posada-Aguirre:2016qpz}. But due to the presence of a partially reflecting surface, the boundary condition near the would-be horizon changes --- from being purely ingoing for the case of black holes --- to harboring both ingoing and outgoing modes, owing to the reflective nature of the surface \cite{Cardoso:2019rvt, Deliyergiyev:2019vti}. These outgoing modes will lead to late-time echoes of the primary signal in the ringdown profile of gravitational waves. Despite having such an interesting structure, which can arise in various different contexts \cite{Dey:2020lhq, Dey:2020pth, Maggio:2018ivz}, there has been one major problem with these models --- one puts forward such a reflective surface close to the horizon in an ad-hoc manner --- neither the reflectivity nor the location of the surface arises from some fundamental point of view. Subsequently, several attempts have been made to model the reflective surface appropriately, either by invoking quantum effects \cite{Maselli:2018fay, Agullo:2020hxe, Chakraborty:2022zlq} or, by considering wormhole geometries \cite{Mark:2017dnq, Cardoso:2008bp, Volkel:2018hwb, PhysRevD.106.124003, Franzin:2022iai}. In both of these scenarios, the existence of such a partially reflective surface is a consequence and not an artifact. Wormholes, in particular, are spacetimes that connect two distinct universes via a throat \cite{visser1995lorentzian, Morris:1988cz}, where these two universes can have their own photon sphere. So that from the perspective of one universe, the angular momentum barrier associated with the photon sphere of the other universe acts as a partially reflecting surface \cite{Mark:2017dnq, PhysRevD.106.124003}. Generic wormhole models have two major drawbacks, first of all the formation channel of wormhole geometries are not known, and secondly, they violate the energy conditions. The second issue, namely the violation of energy conditions, can be circumvented if we consider the braneworld scenario, where the matter field on the brane satisfies the energy conditions, but violations of energy conditions happen in bulk \cite{Kar:2015lma, PhysRevD.106.124003}. The final question regarding these wormhole geometries, and ECOs in general, is their stability. There have been several works discussing the stability of various ECOs \cite{Maggio:2018ivz, Cardoso:2016oxy, Cardoso:2017njb, Zhong:2022jke, Biswas:2021gvq}, including wormhole geometries \cite{PhysRevD.106.124003,Bueno:2017hyj,PhysRevLett.116.171101,DuttaRoy:2022ytr,Bronnikov:2019sbx,Bronnikov:2021liv}, albeit in the isolated scenario. Also, several works have been done to develop astrophysical techniques\cite{DeFalco:2020afv, DeFalco:2021btn, DeFalco:2021klh, DeFalco:2021ksd, DeFalco:2023kqy} to detect wormholes in the extended theories of gravity.
  
However, there are no isolated objects in our universe. As is well known from various experiments --- from the flatness of the rotation curves of galaxies \cite{Rubin1970RotationOT,cowie1986virial, Borriello:2000rv, Persic:1995ru}, the dynamics of hot gas in clusters \cite{Briel:1997hz}, gravitational lensing experiments \cite{SDSS:2005sxd}, among others --- $95$ percent mass of the galaxy comes from that of non-baryonic matter, namely the dark matter \cite{Clowe:2006eq, Freese:2008cz}. Therefore, any compact object, be it a black hole, or, an ECO must reside in an environment involving dark matter, and hence the spacetime geometry must be affected by the same. Recently, a fully relativistic analysis was presented in \cite{Cardoso:2021wlq}, where the spherically symmetric metric of a black hole spacetime was derived in the presence of a galactic matter distribution following the Hernquist-type density profile \cite{hernquist1990analytical}, \footnote{For galactic black hole solutions with other mass profiles see\cite{Konoplya:2022hbl}.}
\begin{equation}\label{Hernquist-density}
\rho(r)= \frac{M a}{2 \pi r (r+a)^3}~,
\end{equation}
where $M$ is the mass of the dark matter halo and $a$ is a typical length scale associated with the dark matter distribution in the galaxy. Motivated by the above density profile, the mass profile of a galactic black hole becomes,
\begin{align}\label{Cardoso-Result-1}
m(r)=M_{\textrm{BH}}+\frac{M r^{2}}{(r+a)^{2}}\left(1-\frac{2M_{\textrm{BH}}}{r}\right)^{2}~, 
\end{align}
where $M_{\textrm{BH}}$ is mass of the central black hole. The geometry resulting from the above mass profile is reminiscent of the Einstein cluster \cite{7bb06a79-8225-31c6-88c3-0c4f8a76b072}. Note that the above mass profile keeps the existence of the black hole horizon intact, even in the presence of a galaxy. It is worthwhile to mention that galactic matter also provides an effective shielding mechanism when the central black hole is electrically charged \cite{Feng:2022evy}. Motivated by the fully relativistic analysis involving galactic black holes, in this paper, we consider the case of a wormhole in the galaxy. Just as galactic black holes keep their horizons intact, we will depict appropriate mass profiles which keep the location of the throat of the galactic wormhole unchanged. We start by considering the environmental effects on the Damour-Solodukhin wormhole and then proceed towards discussing the braneworld wormhole in a galaxy.

In the case of braneworld wormholes one can have two possible of situations --- (i) wormholes in the $\textrm{RS1}$ scenario, involving compact extra dimension and (ii) wormholes in $\textrm{RS2}$ scenario, with non-compact extra dimension. The former case is described by two branes, such that all the standard model particles are localized on the visible brane, while for gravity the effective Einstein's equation takes the form of an scalar-tensor theory of gravity, where the scalar field originates from the inter-brane separation \cite{Kanno:2002ia}. Therefore in this case the scalar field captures any non-local bulk effect. On the other hand, in the latter case the effective Einstein's equation on a vacuum brane inherits an extra contribution from the electric part of the projected bulk Weyl tensor, capturing the non-local bulk effects on the brane. In both the cases discussed above the field equations are not closed in the sense that the solutions of the effective Einstein's equation itself either need the knowledge of the bulk Weyl tensor, or, the evolution of the scslar field. Besides the wormhole geometry in the RS1 model considered here, there are several other wormhole solutions to the gravitational field equations in the context of RS2 scenario as well, with similar as well as distinct solutions \cite{Casadio:2001jg,Bronnikov:2005av,Bronnikov:2019sbx,PhysRevD.67.064027}. It will be worth pursuing the stability of these solutions in the RS2 scenario and subsequently embedding them in a dark matter halo.

Here, we wish to address the question of how the presence of dark matter halo can cure the violation of energy conditions, which were present for the isolated wormhole geometry. Moreover, we will also study the scalar perturbation of galactic wormhole geometries and shall describe the distinguishing features in the ringdown signal of a galactic wormhole in contrast to that of an isolated wormhole spacetime. We also analyze how environmental effects can influence the location of the photon sphere, the ISCO, and finally the shadow radius. 

This paper is organized as follows: In \ref{sec-2} we will discuss the basic strategy to model wormholes involving environmental effects due to the surrounding dark matter halo. Following this in \ref{sec-2.1} we will model a galactic Damour-Solodukhin Wormhole, while  \ref{sec-2.2} will be dedicated to the analysis of the braneworld wormhole. Subsequently, having laid down the geometry of the braneworld wormhole spacetime, in \ref{sec-3} we will study the massless scalar perturbation and shall determine the effective potential experienced by it. Finally, using the transfer matrix method \cite{Bueno:2017hyj} we will find the scalar quasi-normal modes and shall also solve the master equation in the time domain in order to obtain the ringdown waveform. We conclude in \ref{sec-4} with a discussion of our results and provide some future prospects.

\emph{Notations and Conventions:} In our calculations we will set $c=1=G$. The lowercase Greek indices $\mu,\nu,\ldots$ will denote the four-dimensional spacetime coordinates. We will follow the mostly positive signature convention for our metric, such that the Minkowski metric in four spacetime dimensions takes the form $\eta_{\mu \nu}=\textrm{diag.}(-1,+1,+1,+1)$.

\section{Geometry of wormholes at the center of galaxies}\label{sec-2}

In this section, we present the static and spherically symmetric wormhole geometries with a spherical galactic halo surrounding them. By and large, the analysis parallels the case of a galactic black hole but differs in the details and also in the interpretation. Moreover, in the present context, we solve the following Einstein's equations,
\begin{align}\label{Ein_Eq}
G_{\mu \nu}=T_{\mu \nu}^{\rm (g)}+T_{\mu \nu}^{\rm (w)}~,
\end{align}
where, $T_{\mu \nu}^{\rm (g)}$ is the energy-momentum tensor for the galactic matter and $T_{\mu \nu}^{\rm (w)}$ is the energy-momentum tensor supporting the wormhole geometry, often violating the energy conditions. The galactic matter is considered to be distributed spherically around the central wormhole, such that the (wormhole+dark matter halo) can be considered as a static and spherically symmetric configuration together. Moreover, the configuration is supposed to be static, i.e., no radial outflow should be present. Thus, as in the black hole case, here also we let 
\begin{align}
T^{\mu\,\textrm{(g)}}_{\nu}=\textrm{diag.}\left(-\rho^{\rm (g)},0,p_{\perp}^{\rm (g)},p_{\perp}^{\rm (g)}\right)~,
\end{align}
where, $\rho^{\rm(g)}$ is the density of the dark matter halo surrounding the central wormhole and $p_{\perp}^{\rm (g)}$ is the transverse pressure, perpendicular to the radial direction, acting on the dark matter particles. 

The energy-momentum tensor $T_{\mu \nu}^{\rm (w)}$ supporting the wormhole depends on the details of the wormhole geometry and differs from one wormhole solution to another. In what follows, we will discuss the structure of $T_{\mu \nu}^{\rm (w)}$ for two cases --- (a) the Damour-Solodukhin wormhole and (b) the braneworld wormhole. In the case of the Damour-Solodukhin wormhole, $T_{\mu \nu}^{\rm (w)}$ does not arise from some well-defined matter model and is of phenomenological origin, while for braneworld wormhole $T_{\mu \nu}^{\rm (w)}$ depends on the length of the higher dimension and hence have a well-motivated matter model. We will provide explicit expressions for these quantities shortly, which will be useful in constructing the solution of galactic wormholes with an appropriate dark matter profile. 

Since the total configuration involving a wormhole and the galactic matter is spherically symmetric and stationary, for this system, we may consider the following static and spherically symmetric metric ansatz,
\begin{align}\label{sph-metric}
ds^{2}=-f(r)dt^{2}+\left(1-\frac{2m(r)}{r}\right)^{-1}dr^{2}+r^{2}d\Omega_{2}^{2}~,
\end{align}
for describing the geometry of a galactic wormhole. The system has four unknown functions, the mass profile $m(r)$, the $g_{tt}$ component $f(r)$, the energy density $\rho^{\rm (g)}$ and the transverse pressure $p_{\perp}^{\rm (g)}$. While there are three independent equations relating them --- these are the $G^{t}_{t}$ and $G^{r}_{r}$ components of Einstein's equations, and the conservation equation $\nabla_{\mu}T^{\mu}_{\nu}=0$. Here $T^{\mu}_{\nu}=T^{\mu\,\textrm{(w)}}_{\nu}+T^{\mu\,\textrm{(g)}}_{\nu}$ and the covariant derivative is with respect to the full metric \ref{sph-metric}\footnote{In this way we are actually assuming that the two fluids actually do not interact with each other unless gravitationally \cite{Sepulveda:2023pvx,Ciarcelluti:2010ji,Sandin:2008db}.}. 
Thus the system of equations does not close, and hence some additional input is necessary. In the case of black holes, fixing the mass function provides the necessary condition to close the system of equations, which we also follow in the present context involving wormholes. 

The choice of the mass function is constrained from several directions. First of all, we need to choose the mass profile $m(r)$ such that at a large distance from the wormhole throat, it should reproduce the Hernquist-type density profile. While at a small distance, comparable to the wormhole throat, the mass function should resemble that of the wormhole. Having specified the mass profile, one can solve the $G^{r}_{r}$ component of Einstein's equation, as in \ref{Ein_Eq}, to determine the other metric component $g_{tt}$. Finally, the conservation law for the galactic energy-momentum tensor provides the transverse pressure, and the $G^{t}_{t}$ component of Einstein's equations provides the energy density of the dark matter halo. Note that a traversable wormhole requires violation of the null and the weak energy conditions near the throat \cite{visser1995lorentzian} since the wormhole throat acts as a geodesic-defocusing lens. This is achieved by the energy-momentum tensor $T^{\mu\,\textrm{(w)}}_{\nu}$ in \ref{Ein_Eq}, which violates these energy conditions. The above summarizes the basic strategy to solve for the geometry of the galactic wormhole, which we will adopt in the subsequent sections in determining the environmental effect of a galaxy on Damour-Solodukhin and braneworld wormholes.   

\subsection{Damour-Solodukhin wormhole in a galaxy}\label{sec-2.1}

In this section, we provide the environmental effects of galactic dark matter on the Damour-Solodukhin wormhole. For this purpose, we briefly review the isolated \DS wormhole spacetime, which is described by the following line element \cite{Damour:2007ap},
\begin{align}\label{Damour-Iso}
ds^{2}=-\left(1-\frac{2M_{1}}{r}\right)dt^{2}+\left(1-\frac{2M_{2}}{r}\right)^{-1}dr^{2}+r^{2}d\Omega^{2}_{2}~,\qquad~M_{2}=M_{1}(1+\lambda^{2})~.
\end{align}
As evident from the above line element, it follows that $r=2M_{2}$ is a null hypersurface, but it is not a horizon because the norm of the timelike killing vector field $(\partial/\partial t)$ is given by $g_{tt}$, which does not vanish at $r=2M_{2}$. Therefore, timelike observers can remain at rest on that hypersurface, and hence $r=2M_{2}$ is not a Killing horizon, rather is the throat of the wormhole. Moreover, calculation of the Kretschmann scalar $K\equiv R_{\mu \nu \alpha \beta}R^{\mu \nu \alpha \beta}$ shows that $K\varpropto r^{-6}(r-2M_{1})^{-4}$, which is finite at $r=2M_{2}$, but diverges at $r=2M_{1}$. Therefore if one wants to continue the spacetime through the null hypersurface at $r=2M_{2}$, it will have a naked singularity at $r=2M_{1}$, which will violate the weak version of the cosmic censorship conjecture. In order to avoid this pathology, one truncates the spacetime at $r=2M_{2}$ and considers two copies of the spacetime glued at this radius. The resulting spacetime corresponds to the Damour Solodukhin wormhole with a null throat at $r=2M_{2}$. 

The above metric is a solution of Einstein's equations, provided the right-hand side of Einstein's equations involves an energy-momentum tensor with an anisotropic fluid, which has the following components,
\begin{align}\label{Damour-EM-Iso}
T^{\mu\,\textrm{(w)}}_{\nu}=\textrm{diag.}\left(0,p_{r}^{\rm (ds)},p_{\perp}^{\rm (ds)},p_{\perp}^{\rm (ds)}\right)~;
\qquad
p_{r}^{\rm (ds)}=-\frac{\lambda^{2}M_{1}}{4\pi r^{2}(r-2M_{1})}
\qquad
p_{\perp}^{\rm (ds)}=\frac{(r-M_{1})\lambda^{2}M_{1}}{8\pi r^{2}(r-2M_{1})^{2}}~.
\end{align}
As evident from the above energy-momentum tensor, the energy density $\rho^{\rm (ds)}$ supporting the wormhole identically vanishes, while the radial pressure $p_{r}^{\rm (ds)}$ is non-zero but negative. Thus it follows that the wormhole spacetime violates both the null and weak energy conditions, as $\rho^{\rm (ds)}+p^{\rm (ds)}_{r}<0$, while $\rho^{\rm (ds)}+p^{\rm (ds)}_{\perp}>0$. Thus we have described the geometry and the source of the isolated \DS wormhole, which will be used to construct a galactic wormhole with the \DS wormhole at the center.  

\subsubsection{Geometry of the galactic Damour-Solodukhin wormhole}\label{sec-2.1.1}

As already emphasized above, the first step in determining the environmental effect on the \DS wormhole is to provide a suitable mass function with desired properties. Motivated by the corresponding mass profile for galactic black holes, here we consider the following mass profile (for an alternative mass profile, see \ref{app-A}), 
\begin{align}\label{Damour-Gal-Mass-1}
m(r)=M_{2}+\frac{M r^{2}}{(r+a)^{2}}\left(1-\frac{2M_{1}}{r}\right)\left(1-\frac{2M_{2}}{r}\right)~.
\end{align}
Here, $M$ is the mass, and $a$ is the characteristic radius of the dark matter halo. Moreover, there are three cases to consider --- (a) for $r\ll a$, the mass function simply becomes $m(r)\approx M_{2}$, i.e., at radii much smaller compared to the galactic scale $a$, the geometry is that of an isolated \DS wormhole with mass $M_{2}$, (b) for $r\gg M_{2}$, the above mass profile reduces to $M_{2}+\{Mr^{2}/(r+a)^{2}\}$, i.e., the metric is governed by the mass profile of the galactic halo, and finally (c) for $r\gg a$, we obtain for the mass profile $(M_{2}+M)$, which corresponds to the ADM mass of the galactic wormhole spacetime. For the above mass function, the $g^{rr}$ component of the metric describing the galactic wormhole becomes, 
\begin{align}\label{Damour-Gal-grr}
g^{rr}=\left(1-\frac{2M_{2}}{r}\right)\left[1-\frac{Mr}{(r+a)^{2}}\left(1-\frac{2M_{1}}{r}\right)\right]~.
\end{align}
Therefore, $r=2M_{2}$ is still a null hypersurface, and as we will demonstrate later, it also corresponds to the throat of the wormhole. Given the mass profile, and the $g^{rr}$ component of the metric, the $G^{t}_{t}$ component of the Einstein's equations yield, $m'(r)=4\pi r^{2}(\rho^{\rm (g)}_{\rm (ds)}+\rho^{\rm (ds)})$. Since the energy density of matter $\rho^{\rm (ds)}$, supporting the \DS wormhole identically vanishes, for the mass profile in \ref{Damour-Gal-Mass-1}, we obtain the following density profile for the galactic dark matter distribution, 
\begin{align}\label{Damour-gal-density}
\rho^{\rm (g)}_{\rm (ds)}=\frac{2M\left(1-\frac{2M_{1}}{r}\right)(a+2M_{1})}{4\pi r(a+r)^3}
+\lambda^{2}\frac{ M \left[M_{1} (r-4M_{1})-a M_{1}\right]}{2 \pi  r^2 (a+r)^3}~.
\end{align}
Note that the above energy density is non-negative for $r\geq 2M_{2}$. Besides, for $M\rightarrow 0$, the energy density identically vanishes, since it is created by the galactic matter alone. Similarly, for $r\sim 2M_{2}\ll a$, it follows that $\rho^{\rm (g)}_{\rm (ds)}\sim (1/a^{2})$ and vanishes as $a$ becomes large. This is also consistent with the result that the above density is created by the galactic dark matter distribution alone and has no contribution from the energy-momentum tensor for the wormhole. Finally, for $r\gg 2M_{2}$ and for $\lambda\ll 1$, the above density profile reduces to the Hernquist profile, described by \ref{Hernquist-density}, as expected. Similarly one can compute the tangential pressure by solving the $G^{\theta}{}_{\theta}$ component of Einstein's equation or imposing the conservation of energy. The exact expression of the tangential pressure ($p_{\perp}^{(g)}{}_{(ds)}$) is given in the \ref{app-B}.

The determination of the $g_{tt}$ component of the metric is achieved by solving the $G^{r}_{r}$ component of the Einstein's equation, which reads, 
\begin{align}\label{radial-Einstein}
\frac{1}{r}\left(1-\frac{2m(r)}{r}\right)\frac{d\ln f}{dr}-\frac{1}{r^{2}}+\frac{1}{r^{2}}\left(1-\frac{2m(r)}{r}\right)=8\pi \left(p^{\rm (g)}_{r}+p^{\rm (ds)}_{r}\right)~.  
\end{align}
For the galactic matter distribution, there is no radial out/inflows present, and hence we must have $p_{r}^{\rm (g)}=0$. However, unlike the case of black holes, matter distribution supporting wormholes have non-zero radial pressure, which for the \DS wormhole is given by, \ref{Damour-EM-Iso}. For the above radial pressure and the mass profile given by \ref{Damour-Gal-Mass-1}, we find
\begin{align}\label{Damour-gal-gtt}
f(r)=\left(1-\frac{2M_{1}}{r}\right)e^{\gamma}~,
\qquad
\gamma\equiv -\pi \sqrt{\frac{M}{\xi}}+2\sqrt{\frac {M}{\xi}} \tan^{-1}\left(\frac{r+a-M}{\sqrt{M \xi}}\right)~,
\end{align}
where, $\xi\equiv 2a-M+4M_{1}$. It is clear that, like the $g^{rr}$ component, $f(r\rightarrow \infty)\rightarrow 1$ and hence the geometry is asymptotically flat. Interestingly, for the above mass profile, the $g_{tt}$ component is independent of $M_{2}$ and appears in the same form as the galaxy with a central black hole \cite{Cardoso:2021wlq}, provided we make the replacement $M_{1}\rightarrow M_{BH}$. Note that the surface $r=2M_{1}$ is where the norm of the timelike Killing vector field $(\partial/\partial t)$ vanishes, but is not null. Thus the above spacetime, described by the metric elements presented in \ref{Damour-Gal-grr} and \ref{Damour-gal-gtt} depicts a wormhole geometry with its throat located at $r=2M_{2}$, but incorporates the effect of the dark matter halo surrounding the same. We will refer to this geometry as the galactic \DS wormhole. 

\begin{figure}
	\centering
	\minipage{0.5\textwidth}
	\includegraphics[width=\linewidth]{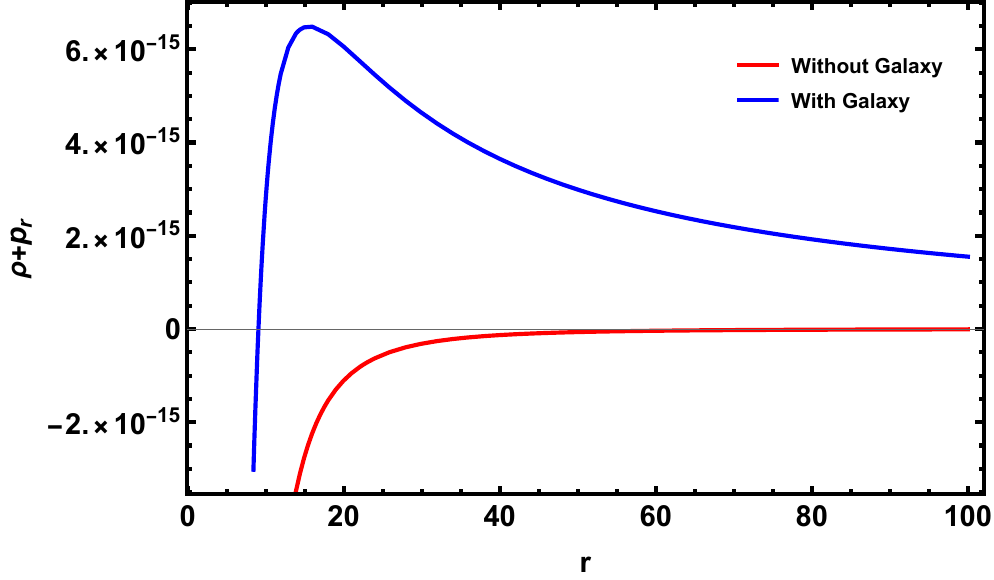}
	\endminipage\hfill
	\minipage{0.5\textwidth}
	\includegraphics[width=\linewidth]{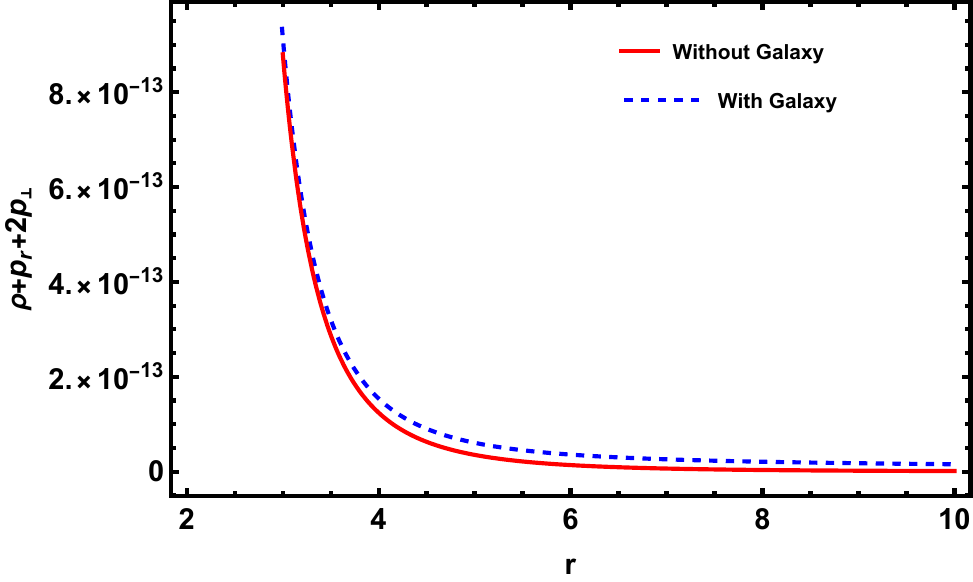}
	\endminipage
	\caption{The quantity $(\rho+p_{r})$ has been depicted on the left as a function of the radial coordinate $r$ for two cases --- (a) without the dark matter halo, and (b) with a dark matter halo, with the following choices of the parameters: $(a/M_{1})=10^{8}$, $(M/M_{1})=10^{4}$, and $\lambda=10^{-5}$. This plot clearly demonstrates that the null energy condition is violated everywhere in the absence of a dark matter halo. But in the presence of a galaxy, i.e., when the wormhole is surrounded by a dark matter halo, the null energy condition is violated in a small radial interval, close to the throat. On the other hand, the right plane plot presents the variation of $\rho+p_{r}+2p_{\perp}$ with the radial coordinate. As evident, this quantity is indeed positive with or without the galaxy.}
	\label{fig-Damour-WEC}
\end{figure}

Let us now explore the status of the energy conditions for the galactic \DS wormhole, in particular, it follows that total matter-energy density supporting this wormhole is simply given by $\rho^{\rm (g)}_{\rm (ds)}$, which is positive for $r>2M_{2}$, and hence the weak energy condition is satisfied everywhere outside the wormhole. To see what happens for the null energy condition, we have plotted the sum of total energy density and radial pressure against the radial coordinate $r$ in \ref{fig-Damour-WEC}. As evident, the energy density comes from the galactic dark matter, while the radial pressure arises from the matter supporting the wormhole, such that $\rho+p_{r}\equiv\rho^{\rm (g)}_{\rm (ds)}+p_{r}^{\rm (ds)}$, among which the contribution by the energy density is positive, while the radial pressure contributes negatively. As a consequence, and as clear from \ref{fig-Damour-WEC}, the violation of the null energy condition for the galactic \DS wormhole happens for a much smaller spatial region ($2M_{2}\leq r<r_{0}$), compared to the \DS wormhole, where the violation is present throughout the space ($2M_{2}\leq r<\infty$). The radius $r_{0}$, below which the violation of the null energy condition happens can be obtained by solving the algebraic equation $\rho+p_{r}=0$, which yields,
\begin{align}\label{zero-crossing}
r_{0}\simeq 2M_{2}+a\sqrt{\frac{M_{2}}{2M}} \lambda~,
\end{align}
definitely greater than the throat location of the wormhole. Thus it is appropriate to argue that in the presence of a galactic halo, the violation of the null energy condition can be tamed, such that for all space outside the radial coordinate $r_{0}$ the null energy condition is satisfied\footnote{One can also show that $\rho+p_{\perp}$ is positive, as well as $\rho+p_{r}+2p_{\perp}>0$, for $r\geq 2M_{2}$. Thus strong energy condition is also satisfied everywhere in the galactic \DS wormhole for $r\geq r_{0}$.}. 

\begin{figure}
\centering
\minipage{0.5\textwidth}
\includegraphics[width=\linewidth]{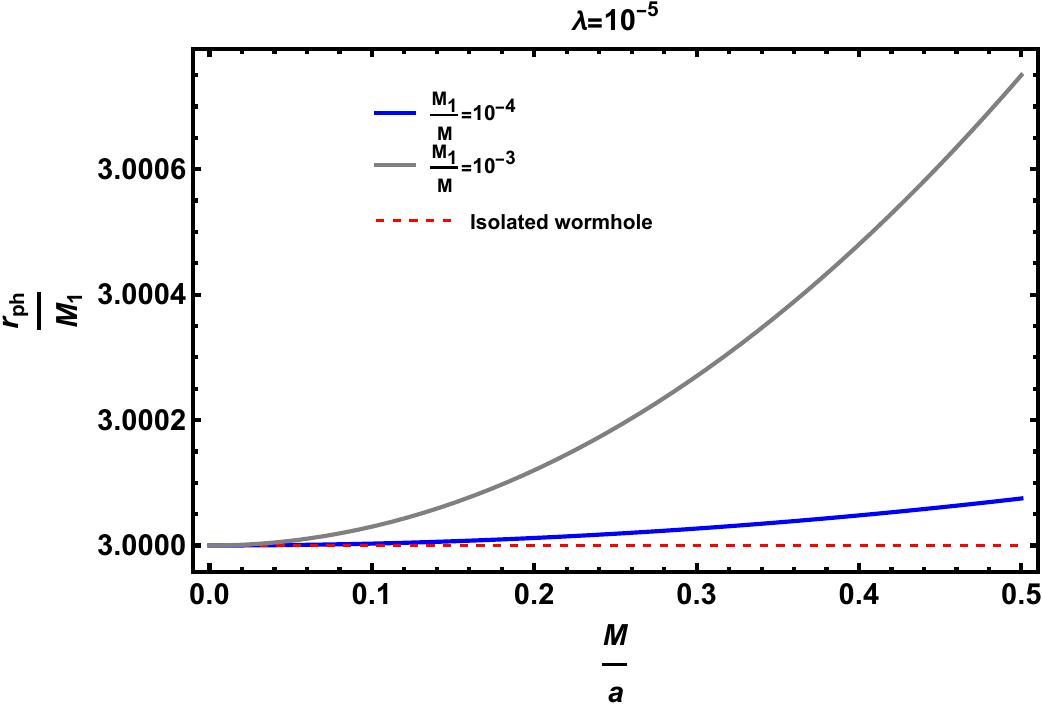}
\endminipage\hfill
\minipage{0.47\textwidth}
\includegraphics[width=\linewidth]{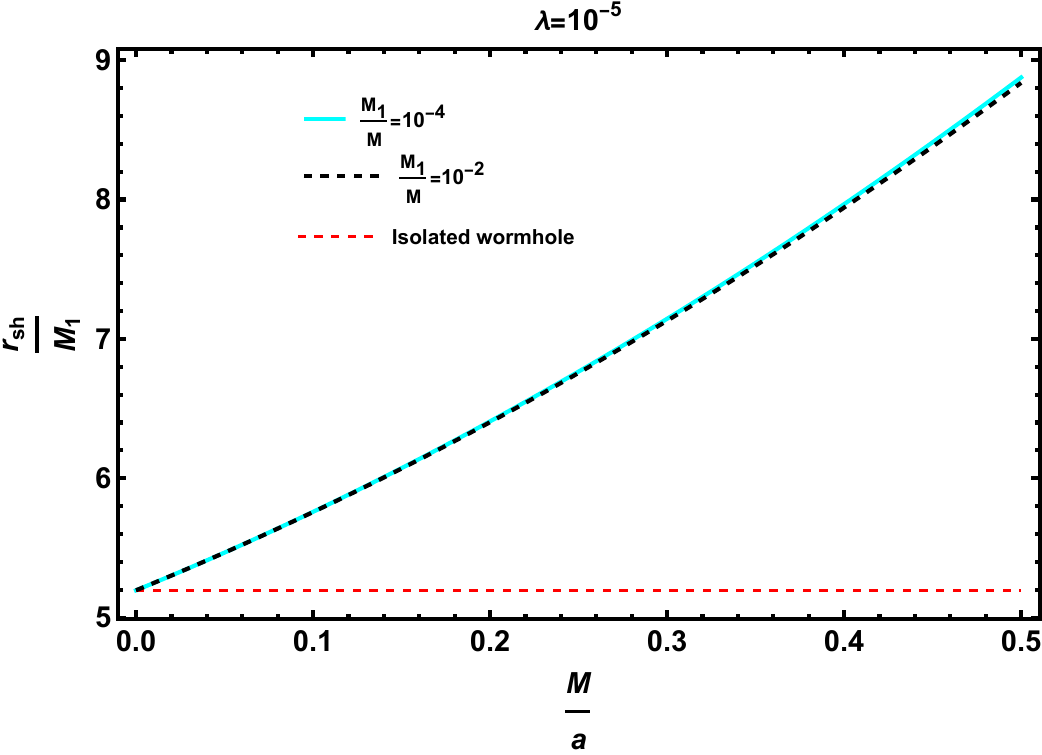}
\endminipage
\caption{The radial location of the photon sphere $r_{\rm ph}$ as a function of the compactness of the galaxy ($M/a$) has been presented in the left. The plot shows that in the presence of a galaxy, the radial location of the photon sphere increases compared to that of the isolated \DS wormhole. On the other hand, the right panel demonstrates the variation of the shadow radius $r_{\rm sh}$ of galactic \DS wormhole as a function of the galactic compactness ($M/a$) and depicts a similar behaviour. All the radii are scaled by the ADM mass of the wormhole $M_{1}$.}\label{fig-Damour-ph-sh}
\end{figure}

\subsubsection{Properties of the galactic Damour-Solodukhin wormhole}\label{sec-2.1.2}

At this stage, it will be instructive to specify the hierarchy between various mass scales and $\lambda$. To mimic galactic observations, we assume that $a>M\gg M_{1}$\cite{Navarro:1995iw} and restrict the parameter $\lambda$, such that $\lambda \ll 1$. The first assumption simply states that the size of the galactic dark matter distribution is much larger than the size of the wormhole. This is typical of any astrophysical scenario, where the size of the central compact object is several orders of magnitude smaller than the size of the galaxy. The second assumption makes the underlying wormhole spacetime, described by \ref{Damour-Iso} close to that of Schwarzschild black hole mimicker. Using these assumptions, we can expand the relevant expressions in the powers of $\lambda$ and the dimensionless ratio $(M_{1}/a)$. Such expansion enables us to determine the closed-form expressions for various quantities associated with the galactic \DS wormhole, e.g., the radius of the photon sphere and the innermost stable circular orbit (in short, ISCO). 

In what follows we will provide a brief derivation of the equation governing the location of the photon sphere. Since photons move along null geodesics, we can take $l^{\mu}$ to be the tangent along the null geodesics, satisfying $l^{\mu}l_{\mu}=0$. Due to the Killing symmetries of the spacetime we have conserved energy $E=-l_{\mu}(\partial/\partial t)^{\mu}$ and conserved angular momentum $L=l_{\mu}(\partial /\partial \phi)^{\mu}$. The condition, $l^{\mu}l_{\mu}=0$, yields the following equation for the radial component of the null geodesic, 
\begin{align}\label{null}
g_{rr}(l^{r})^{2}-\frac{L^{2}}{r^{2}}=-g^{tt}\frac{L^{2}}{b^{2}}~,
\end{align}
where the impact parameter is defined as $b=(L/E)$. It is evident from \ref{null} that the potential experienced by any null geodesics, including photon, is $V_{\textrm{ph}}=-(L^{2}/r^{2})g_{tt}$. Using the conditions of having a circular photon orbit --- (a) $V_{\textrm{ph}}=0$, and (b) $V^{\prime}_{\textrm{ph}}=0$, leads to the condition $rg_{tt}^{\prime}(r)=2g_{tt}$ at $r=r_{\textrm{ph}}$. Solving this algebraic equation in the present context yields the following equation for the photon sphere, to leading order in $(M_{1}/a)$,
\begin{align}\label{Damour-Photon}
r_{\rm ph}=3M_{1}\left(1+\frac{MM_{1}}{a^{2}}\right)~.
\end{align}
It is to be emphasized that the photon sphere as obtained above is a special case of photon region \cite{Galtsov:2019fzq,Kato:2023qxq,Kobialko:2020vqf,Bogush:2021qnz,Kobialko:2021uwy,yassine2022flat}, which is defined as a compact region of the spacetime where photons can travel endlessly without going to infinity or to the event horizon.

Note that in the limit of $M\rightarrow 0$, from \ref{Damour-Photon} we get back the result for the photon sphere of the isolated \DS wormhole, as we should. Also, the presence of a dark matter halo increases the radial location of the photon sphere, compared to the case of an isolated wormhole. The computation of the ISCO radius follows from the effective potential experienced by a massive particle moving in this spacetime, in particular, one simultaneously imposes three conditions: $V_{\rm eff}=0$, $V_{\rm eff}'=0$ and $V_{\rm eff}''=0$, which provides an algebraic equation for the radial coordinate with the following solution,   
\begin{align}\label{Damour-ISCO}
r_{\rm ISCO}=6M_{1}\left(1-\frac{32M M_{1}}{a^{2}}\right)~.
\end{align}
Here also, in the limit of $M\rightarrow 0$, we get back the ISCO radius associated with an isolated \DS wormhole, however, in contrast to the photon sphere, the radial location of ISCO for the galactic \DS wormhole decreases compared to the isolated \DS wormhole. Along similar lines, one can compute the shadow radius associated with the galactic \DS wormhole, which is given by
\begin{align}\label{Damour-Shadow}
r_{\rm sh}=3\sqrt{3}M_{1}\left[1+\frac{M}{a}+\frac{M(5M-18M_{1})}{6 a^{2}}\right]~.
\end{align}
As for the case of the photon sphere, the shadow radius also reduces to the desired expression $3\sqrt{3}M_{1}$ in the absence of galactic matter, i.e., in the limit $M\rightarrow 0$, and increases in the presence of galactic matter (as $M\gg M_{1}$). This suggests that the shadow radius of any compact object in a galactic environment will be larger compared to the isolated object. It is worth pointing out that all the quantities derived above, namely the photon sphere $r_{\rm ph}$, the ISCO radius $r_{\rm ISCO}$ and the shadow radius $r_{\rm sh}$ depend on the $g_{tt}$ component and its derivatives alone, and hence these are all independent of the wormhole parameter $\lambda$. Both the photon sphere and the shadow radii are plotted in \ref{fig-Damour-ph-sh}, against the dimensionless galactic scale $(M/a)$. The figure explicitly demonstrates how the photon sphere and the shadow radius are affected by the presence of a galaxy, in particular, it follows that the presence of dark matter halo increases the radii compared to the isolated wormhole counterpart.

\subsection{Galactic wormhole in the braneworld scenario}\label{sec-2.2}

We start by briefly summarizing the isolated braneworld wormhole scenario. In this case, one assumes that we live in a four-dimensional hypersurface, known as the brane, embedded in a five-dimensional spacetime, referred to as the bulk. The bulk spacetime is generically taken to be endowed with a negative cosmological constant, while the extra spacelike extra dimension (denoted as $y$) is taken to be compact. We consider a two-brane system, where the Planck brane is located at $y=0$ (typical energy scale in this brane is Planckian order) and the visible brane is located at $y=\ell$ (energy scale in this brane is a few tens of TeV). The metric in the bulk spacetime can be expressed as \cite{Kanno:2002ia},
\begin{align}\label{brane-metric-ansatz}
ds^{2}=e^{2\phi(x)}dy^{2}+\widetilde{g}_{\mu\nu}(y,x)dx^{\mu}dx^{\nu}~,
\end{align}
where $\phi(x)$ is called the radion field and is related to the separation between the two branes, such that the proper distance between the two branes is given by $d(x)=e^{\phi(x)}\ell$. It is expected that the curvature length scale of the bulk spacetime, which is given by $\ell$, will be much smaller compared to the corresponding length scale on the brane. This is because the curvature of the bulk spacetime is much larger than the curvature on the brane, and hence an inverse relationship exists between the respective length scales. Thus, we can expand the gravitational field equations as a power series in ($\ell /\textrm{brane~curvature~scale})$. This leads to the following effective Einstein's equation on the visible brane \cite{Kanno:2002ia},
\begin{align}\label{Visible-eom-B}
G_{\mu\nu}=\frac{\kappa^{2}}{\ell \Phi}T^{\rm (vis)}_{\mu\nu}+\frac{\kappa^{2}(1+\Phi)}{\ell \Phi}T^{\rm (Pl)}_{\mu\nu}+\frac{1}{\Phi}T^{\Phi}_{\mu\nu}~;
\qquad
\Phi=\exp [2e^{\phi(x)}]-1~,
\end{align}
where, 
\begin{align}\label{energy-Phi}
T^{\Phi}_{\mu\nu}=\bigg(\nabla_{\mu}\nabla_{\nu}\Phi-g_{\mu\nu}\nabla^{\alpha}\nabla_{\alpha}\Phi\bigg)-\frac{3}{2(1+\Phi)}\bigg(\nabla_{\mu}\Phi\nabla_{\nu}\Phi-\frac{1}{2}g_{\mu\nu}\nabla_{\alpha}\Phi\nabla^{\alpha}\Phi\bigg)~.
\end{align}
In the above expression for the effective gravitational field equations on the visible brane, presented in \ref{Visible-eom-B}, except for the energy-momentum tensor $T^{\Phi}_{\mu\nu}$, arising from the radion field, we have two additional contributions --- $T^{\rm (vis)}_{\mu\nu}$ and $T^{\rm (Pl)}_{\mu\nu}$, which are the energy-momentum tensors of the matter fields trapped on the visible brane and the Planck brane, respectively. Moreover, the ratio $(\kappa^{2}/\ell)$ acts as an effective gravitational constant on the brane, where $\kappa^{2}$ is the bulk gravitational constant. Note that the metric $g_{\mu \nu}$ on the visible brane is related to the metric $\widetilde{g}_{\mu \nu}$, appearing in \ref{brane-metric-ansatz}, by a conformal factor $\exp(-d/\ell)$. Besides the gravitational field equations on the brane, we also obtain a field equation for the radion field $\Phi$, which reads \cite{Kanno:2002ia},
\begin{align}\label{Brane-EOM-Phi}
\nabla^{\alpha}\nabla_{\alpha}\Phi=\frac{\kappa^{2}}{\ell}\frac{T^{\rm (Pl)}+T^{\rm (vis)}}{2\omega+3}-\frac{1}{2\omega+3}\frac{d\omega}{d\Phi}(\nabla^{\alpha}\Phi)(\nabla_{\alpha}\Phi)~\quad\text{with}\quad\omega=-\frac{3\Phi}{2(1+\Phi)}~,
\end{align}
where $T^{\rm (vis)}$ and $T^{\rm (Pl)}$ are, respectively, the traces of the matter energy-momentum tensor of the visible and the Planck brane. In what follows, we will consider the case of a vacuum Planck brane, i.e., we will assume $T^{\rm (Pl)}_{\mu\nu}=0$. Note that, the Bianchi identity when applied to \ref{Visible-eom-B} must imply the field equation for the radion field for consistency, which we have reproduced in \ref{app-C}. Given the gravitational field equations on the brane and the radion field equations presented above, in the context of static and spherically symmetry, the following solution can be obtained (for details, see \cite{Kar:2015lma}),
\begin{align}\label{Metric-brane-sol}
&ds^{2}=-\frac{1}{(\chi+1)^{2}}\left(\chi+\sqrt{1-\frac{2M_{1}}{r}}\right)^{2}dt^{2}+\left(1-\frac{2M_{1}}{r}\right)^{-1}dr^{2}+r^{2}d\Omega_{2}^{2}~,
\end{align}
with the radial dependence of the radion incarnation $\Phi$ on the visible brane being,
\begin{align}\label{radion-iso-in-r}
\Phi(r)=\frac{\Phi^{2}_{1}}{4}\left[\ln\left(\chi+\sqrt{1-\frac{2M_{1}}{r}}\right)\right]^{2}+\Phi_{1}\sqrt{\Phi_{0}+1}\ln\left(\chi+\sqrt{1-\frac{2M_{1}}{r}}\right)+\Phi_{0}~,   
\end{align} 
here  $\Phi_{1}$ and $\Phi_{0}$  are some constants and chosen in such a way that the two branes never collide \cite{PhysRevD.106.124003}. Note that the field equation for $\Phi$ admits two solutions, the above solution is valid for $\chi\neq 0$, while for $\chi=0$, we have $\Phi=\textrm{constant}$. Therefore, non-zero values of the parameter $\chi$ yield a non-trivial scalar field configuration and also a distinct static and spherically symmetric geometry. For $\chi=0$, we get back the Schwarzschild solution with a constant radion field $\Phi$, as expected. The above solution for the field $\Phi$ can also be expressed in terms of isotropic coordinate $r'$, related to the radial coordinate $r$ by the following relation: $r=r^{\prime}\left\{1+(M_{1}/2r^{\prime})\right\}^{2}$, such that \ref{radion-iso-in-r} reduces to the corresponding expression presented in \cite{Kar:2015lma}. For our purpose, the solution for $\Phi$ in terms of the radial coordinate $r$ will be sufficient. 

The above solution for the metric $g_{\mu \nu}$ and the field $\Phi$ arises based on the assumption that the Ricci scalar of the visible brane vanishes and the brane matter is an anisotropic fluid, described by the following energy-momentum tensor $T^{\mu\,\textrm{(vis)}}_{\nu}=\textrm{diag}(-\rho^{\rm (vis)},p^{\rm (vis)}_{r},p^{\rm (vis)}_{\perp},p^{\rm (vis)}_{\perp})$ such that its trace vanishes \cite{Dadhich:2001fu, PhysRevD.65.084040, PhysRevD.67.064027}. However, what enters into the gravitational field equations is a combination of the brane energy-momentum tensor $T^{\textrm{(vis)}}_{\mu \nu}$ and the radion field energy-momentum tensor $T^{\Phi}_{\mu \nu}$. Thus, in our subsequent calculations, we will only need the expressions for the components of the total energy-momentum tensor, which reads, 
\begin{align}\label{Brane-EM-Iso}
T^{\mu\, \textrm{(b)}}_{\nu}&\equiv\frac{1}{\Phi}T^{\mu\,\textrm{(vis)}}_{\nu}+\frac{\ell}{\kappa^{2}\Phi}T^{\mu\,\Phi}_{\nu}
=\textrm{diag.}\left(0,p^{\rm (b)}_{r},p^{\rm (b)}_{\perp},p^{\rm (b)}_{\perp}\right)~;
\nonumber
\\
p^{\rm (b)}_{r}&=-\frac{ \chi M_{1} }{8 \pi r^3   \left(\chi+\sqrt{1-\frac{2 M_{1}}{r}}\right)}~,
\qquad
p^{\rm (b)}_{\perp}=\frac{2 \chi M_{1} }{8 \pi r^3   \left(\chi+\sqrt{1-\frac{2 M_{1}}{r}}\right)}~.
\end{align}
Note that for $\chi=0$, the line element in \ref{Metric-brane-sol}, reduces to that of Schwarzschild, and $\Phi$ becomes constant, therefore non-dynamical. For $\chi>0$, \ref{Metric-brane-sol} represents a wormhole solution with a null throat at $r=2M_{1}$, this is because at this radius the metric component $g^{rr}$ vanishes, but the time-like killing vector field $(\partial/\partial t)^{\mu}$, remains timelike. Thus, the surface $r=2M_{1}$ is not a Killing horizon; hence, the above spacetime does not describe a black hole geometry. On the other hand, the spacetime becomes complex beyond $r=2M_{1}$, thanks to the presence of the square root term, and hence is not extendable beyond this radius. Thus one joins two copies of the spacetime at $r=2M_{1}$, akin to wormhole geometries, and this radius is referred to as the throat of the wormhole. Therefore, \ref{Metric-brane-sol} describes a wormhole geometry, mimicking the Schwarzschild spacetime for small wormhole parameter $\chi$ and large $r$, with $\chi\gtrsim~0$. Moreover, the total energy-momentum tensor $T^{\rm (b)}_{\mu \nu}$ violates the weak as well as the null energy conditions everywhere, as $\rho^{\rm (b)}+p_{r}^{\rm (b)}<0$, which is another characteristic feature of wormhole geometries. 

\subsubsection{Geomtry of the galactic braneworld wormhole}\label{sec-2.2.1}

Having outlined the geometry and the associated gravitational field equations of an isolated wormhole spacetime on the brane, we now incorporate environmental effects due to the surrounding dark matter halo. The geometry is taken to be static and spherically symmetric, such that the line element of the galactic braneworld wormhole is still given by \ref{sph-metric}. However, the method adopted in the context of \DS wormhole does not apply here due to the non-polynomial nature of the background metric describing the braneworld wormhole, in particular, the square root term in the $g_{tt}$ component of the metric. Also, Einstein's equations in the presence of galactic matter on the brane become, 
\begin{align}\label{Brane-gal-Einstein}
G_{\mu \nu}=\frac{\kappa^{2}}{\ell}\left(\frac{1}{\Phi}T_{\mu \nu}^{\rm (g)}+T_{\mu \nu}^{\rm (b)}\right)~,
\end{align}
where $T_{\mu \nu}^{\rm (b)}$ contains the contribution from the scalar and the matter on the visible brane, which for the isolated wormhole case has already been presented in \ref{Brane-EM-Iso} and the energy-momentum tensor of the galactic matter is taken to be $T^{\mu\,\textrm{(g)}}_{\nu}=\textrm{diag.}(-\rho^{\rm (g)},0,p_{\perp}^{\rm (g)},p_{\perp}^{\rm (g)})$. Since we have modeled the galaxy using an anisotropic fluid with vanishing radial pressure, in the presence of the galaxy we have five unknowns --- (i) $\rho^{\rm (g)}$, (ii) $p_{\perp}^{\rm (g)}$, (iii) $g_{tt}$, (iv) $g_{rr}$, and (v) $\Phi$. On the other hand, we have three equations at our disposal, namely (a) two Einstein's equations corresponding to $G^{r}{}_{r}$ and $G^{t}{}_{t}$ components, (b) the field equation for the radion incarnation $\Phi$. Note that the conservation of the energy-momentum tensor follows from the above three equations. Therefore, to close the system of differential equations and in order to determine the radial dependence of the field $\Phi$, we fix the metric component $g_{tt}$ and the mass profile $m(r)$ such that,
\begin{align}
f(r)&=\frac{1}{(\chi+1)^{2}}\left(\chi+\sqrt{1-\frac{2M_{1}}{r}}\right)^{2} e^{\gamma}~;
\quad 
\gamma=-\pi \sqrt{\frac{M}{\xi}}+2\sqrt{\frac {M}{\xi}}\tan^{-1}\left[\frac{r+a-M}{\sqrt{M \xi}}\right]~,
\label{brane-gtt-fix}
\\
m(r)&= M_{1}+\frac{M r^{2}}{(r+a)^{2}}\left(1-\frac{2M_{1}}{r}\right)^2~,
\label{Brane-Mass-ansatz}
\end{align}
with, $\xi=2a-M+4M_{1}$.  Here $M$ is the mass of the dark matter halo, and $a$ is a typical galactic length scale. The above choice is motivated by several results --- (a) In the limit $\chi\rightarrow 0$, the above expression for $g_{tt}$ reduces to the solution for a galactic black hole, presented in \cite{Cardoso:2021wlq}, as it should; (b) In the limit $M\rightarrow 0$, the function $f(r)$ reduces to the corresponding metric component of the isolated braneworld wormhole, and (c) the Killing vector $(\partial/\partial t)$ remained timelike at $r=2M_{1}$, typical of a wormhole spacetime. Along identical lines, it also follows that the mass profile has all the desired properties as described above, and in addition, we have --- (a) the surface $r=2M_{1}$ is a null hypersurface, since $g^{rr}$ vanishes there, and (b) for $r\gg 2M_{1}$, the above mass profile reduces to the Hernquist profile. Therefore the above is also a viable mass profile for the braneworld wormhole embedded in a dark matter halo. 

Using these metric components, we can calculate the Ricci scalar associated with the galactic braneworld wormhole, while keeping terms up to $\mathcal{O}(1/a^{2})$, which reads,
\begin{align}\label{Ricci-gal-brane}
R=\frac{2 M \left(12 M_{1}^2-10 M_{1} r+2 r^2\right)}{a^2 r^2 (r-2 M_{1})}~.
\end{align}
As expected, the Ricci scalar $R$ vanishes for $M\rightarrow 0$, i.e., for the isolated wormhole. In particular, taking the trace of \ref{Brane-gal-Einstein}, and using the result that $T^{\mu\,\textrm{(vis)}}{}_{\mu}=0$, as well as the equation of motion of the radion incarnation field $\Phi$,
\begin{align}\label{Phi-EOM-Explicit}
\nabla^{\alpha}\nabla_{\alpha}\Phi-\frac{1}{2(1+\Phi)}\nabla^{\alpha}\Phi\nabla_{\alpha}\Phi=\frac{\kappa^{2}}{\ell}\frac{(1+\Phi)T^{(g)}}{3}~,
\end{align}
it follows that, 
\begin{align}\label{R-via-T-gal}
R=-\frac{\kappa^{2}}{\ell}T^{(g)}~.
\end{align}
This result shows that the Ricci scalar for the galactic braneworld wormhole only depends on the trace of the galactic matter. Further, using \ref{Ricci-gal-brane} we can write the above expression as,
\begin{align}\label{rho-pt-gal}
\rho^{(g)}-2p^{(g)}_{\perp}=\frac{\ell}{\kappa^{2}}\frac{2 M \left(12 M_{1}^2-10 M_{1} r+2 r^2\right)}{a^2 r^2 (r-2 M_{1})}~.
\end{align}
To determine the explicit from of $p^{(g)}_{\perp}$ and $\rho^{(g)}$ one needs to know the solution for the field $\Phi$, which can be obtained by solving \ref{Phi-EOM-Explicit}. However, the non-polynomial nature of the metric components poses a serious problem for obtaining the most general analytic solution of $\Phi$. Thus, we assume that $\Phi$ depends on only radial coordinate and hence express the field equation, namely \ref{Phi-EOM-Explicit}, in the following form,
\begin{align}\label{Phi-to-epsilon}
\frac{d^{2}\epsilon}{dr^{2}}+\frac{d}{dr}\left[\ln{r^{2}\sqrt{f(r)\left(1-\frac{2m(r)}{r}\right)}}\right] \frac{d\epsilon}{dr}=-\frac{\epsilon}{6}R\left(1-\frac{2m(r)}{r}\right)~,\quad \Phi(r)\equiv \epsilon^{2}(r)-1~. 
\end{align}
The right-hand side of the above equation, using \ref{Ricci-gal-brane}, can be shown to be $\sim \{M(r-2M_{1})/a^{2}\}$, which vanishes near the wormhole throat located at $r=2M_{1}$. Thus it follows that near the throat the right-hand side of \ref{Phi-to-epsilon} is ignorable, while the radion incarnation $\Phi$ is only dominant near the throat. Then we can ignore the right hand side term in \ref{Phi-to-epsilon} and get the following approximate solution for $\Phi(r)$,
\begin{align}\label{Phi-brane-gal-sol}
\Phi(r)&\approx \frac{\Phi^{2}_{1}}{4}\left(1-\frac{4 M M_{1}}{a^{2}}\right)\ln\left(\chi+\sqrt{1-\frac{2M_{1}}{r}}\right)^{2}
\nonumber
\\
&\qquad +\Phi_{1}\sqrt{\Phi_{0}+1}\left(1-\frac{2 M M_{1}}{a^{2}}\right)\ln\left(\chi+\sqrt{1-\frac{2M_{1}}{r}}\right)+\Phi_{0}~.
\end{align}
As evident, structurally, the field $\Phi$ in the context of the galactic braneworld wormhole is identical to the isolated wormhole scenario, but a comparison with \ref{radion-iso-in-r} reveals that in the presence of galactic matter, the field $\Phi$ is simply getting re-scaled.

\begin{figure}
\centering
\minipage{0.5\textwidth}
\includegraphics[width=\linewidth]{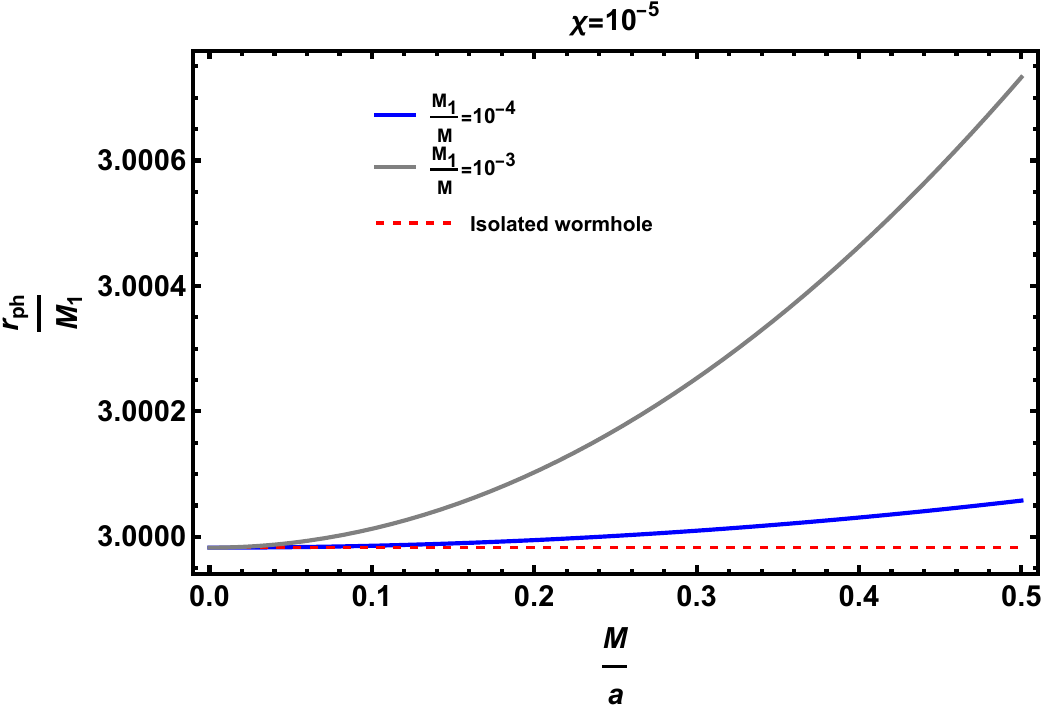}
\endminipage\hfill
\minipage{0.47\textwidth}
\includegraphics[width=\linewidth]{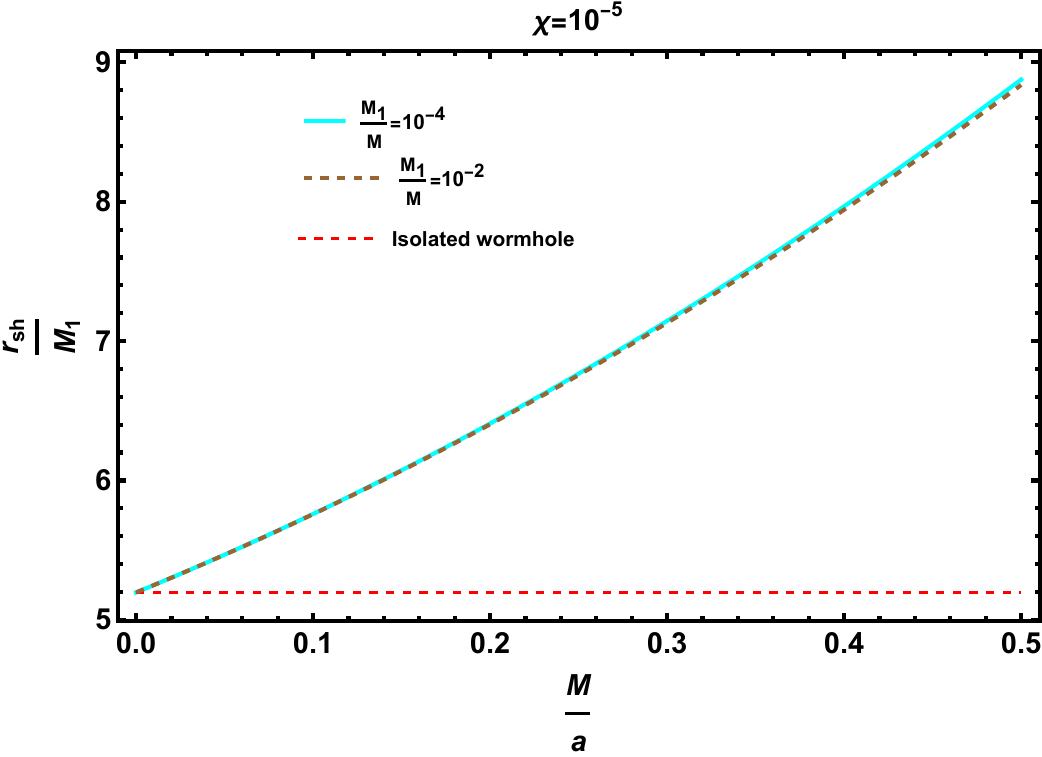}
\endminipage
\caption{The radial location of the photon sphere $r_{\rm ph}$ has been presented as a function of the compactness of the galaxy described by the ratio ($M/a$) on the left. The plot shows that in the presence of a galaxy, the radial location of the photon sphere increases compared to that of the isolated wormhole. On the other hand, the right panel demonstrates the variation of the shadow radius $r_{\rm sh}$ of galactic braneworld wormhole as a function of the galactic compactness ($M/a$). This plot also depicts a similar behaviour, i.e., in the presence of a galaxy the shadow radius increases compared to its isolated counterpart. In both cases all the radii are scaled by the ADM mass of the wormhole $M_{1}$.}\label{fig-Brane-ph-sh}

\end{figure}

\subsubsection{Properties of the galactic braneworld wormhole}\label{sec-2.2.2}

The above discussion provides a closure to the discussion involving the geometry of the galactic braneworld wormhole. The detailed expressions for the energy density $\rho^{\rm (g)}$ and transverse pressure $p_{\perp}^{\rm (g)}$ can also be obtained from Einstein's equations, which we will not present here due to the complicated nature of the corresponding expressions. Before concluding this section, we present below the shadow radius and the radius of the ISCO for the galactic braneworld wormhole and compare it with the isolated case. As stated earlier, to mimic galactic observation, we will restrict ourselves in the parameter space such that $a>M>>M_{1}$ and $\chi\gtrsim 0$, so that we keep terms up to $\mathcal{O}(1/a^{2})$ and linear order in $\chi$, in what follows. Using this approximation, the radii of the photon sphere and the ISCO are given by,
\begin{align}\label{brane-ph-ISCO}
r_{\rm ph}&=3 M_{1} \left(1+\frac{M M_{1}}{a^2}\right)-\sqrt{3} \chi M_{1}~, 
\\
r_{\rm ISCO}&=6 M_{1} \left(1-\frac{32 M M_{1}}{a^2}\right)-3 \sqrt{\frac{3}{2}} \chi M_{1}~.
\end{align}
The presence of galactic matter places the photon sphere $r_{\rm ph}$ at a larger radius, while the exotic material at the throat decreases the position of the photon sphere. Therefore, a competition between galactic matter and wormhole matter exists, and the photon sphere of a galactic braneworld wormhole is at a larger radius than the isolated wormhole counterpart. From the above expression of $r_{\rm ISCO}$, it is clear that in the presence of a galaxy, the ISCO location decreases compared to the isolated wormhole counterpart. Similarly, one can compute the location of the shadow radius for the galactic braneworld wormhole, given by,
\begin{align}\label{Brane-Shadow}
r_{\rm sh}=3\sqrt{3}M_{1}\left[\left(1+\frac{M}{a}\right)\left(1+(1-\sqrt{3})\chi\right)+\frac{M(5M-18M_{1})}{6 a^{2}}\right]~.
\end{align}
Both the photon sphere and the shadow radii are plotted in \ref{fig-Brane-ph-sh}, against the dimensionless galactic scale $(M/a)$. The figure explicitly demonstrates how the photon sphere and the shadow radius are affected by the presence of a galaxy, in particular, it follows that the presence of dark matter halo increases the radii compared to the isolated wormhole counterpart.

\section{Stability of galactic wormholes under scalar perturbation}\label{sec-3}

In this section, we will study the stability of the galactic wormhole spacetimes derived in the previous sections, under external scalar perturbation. In particular, we wish to study how a massless scalar field $\Psi$ evolves in the background geometry, presented in \ref{sec-2.1} and \ref{sec-2.2}, with appropriate mass profiles for the dark matter environment in the context of \DS wormhole, as well as in the braneworld scenario, respectively. We will assume that the energy density of the scalar field $\Psi$ is small, such that it can be ignored compared to the energy densities of various matter species appearing in the galactic wormhole spacetimes. Thus the scalar field $\Psi$ can be considered as a test scalar field, living in the background geometry of the galactic wormholes, and satisfying the massless Klein-Gordon equation, $g^{\mu\nu}\nabla_{\mu}\nabla_{\nu}\Psi=0$. For the metric $g_{\mu \nu}$, we can exploit the result that it describes a static and spherically symmetric geometry, such that the scalar field admits the following decomposition,
\begin{align}\label{Damour-Scalar-decom}
\Psi(t,r,\theta,\phi)=\frac{1}{r}\sum_{l=0}^{\infty}\sum_{m=-l}^{l}e^{-i\omega t}Y_{lm}(\theta,\phi)\psi_{lm}(r)~,
\end{align}
where, $Y_{lm}(\theta,\phi)$ corresponds to the spherical harmonics and $\psi_{lm}(r)$ is the radial function which needs to be determined. Till this point, we have not used any explicit form for the metric, describing the background spacetime. Substituting the above decomposition for the scalar field $\Psi$ in the massless Klein-Gordon equation, along with the metric ansatz from \ref{sph-metric}, we obtain the following equation satisfied by the radial part $\psi_{lm}$ of the scalar field,
\begin{align}\label{Damour-Radial-Master}
\frac{d^{2}\psi_{lm}}{dr_{*}^{2}}+\Big[\omega^{2}-V_{l}(r)\Big]\psi_{lm}=0~.
\end{align} 
Here we have defined the tortoise coordinate $r_{*}$ as the solution of the following differential equation, 
\begin{align}\label{tortoise}
\frac{dr_{*}}{dr}=\frac{1}{\sqrt{g^{rr}g_{tt}}}=\sqrt{\frac{r}{f(r)\left[r-2m(r)\right]}}~,
\end{align} 
and the potential $V_{l}(r)$ is given by, 
\begin{align}\label{gen_pot}
V_{l}(r)= f (r) \frac{l(l+1)}{r^2}+ \frac{\sqrt{f(r)\left[1-2m(r)/r\right]}}{r}\partial_{r}\left(\sqrt{f(r)\left[1-2m(r)/r\right]}\right)~.
\end{align} 
To proceed further and to obtain the characteristic quasi-normal mode frequencies of the galactic wormhole spacetimes, we need to provide explicit expressions for the mass function and the function $f(r)$, which depends on the specifics of the solution considered and will differ from the galactic \DS wormhole to the galactic braneworld wormhole. Thus in what follows, we present the stability analysis of these two galactic wormhole spacetimes separately. 

\begin{figure}
	\centering
	\minipage{0.7\textwidth}
	\includegraphics[width=\linewidth]{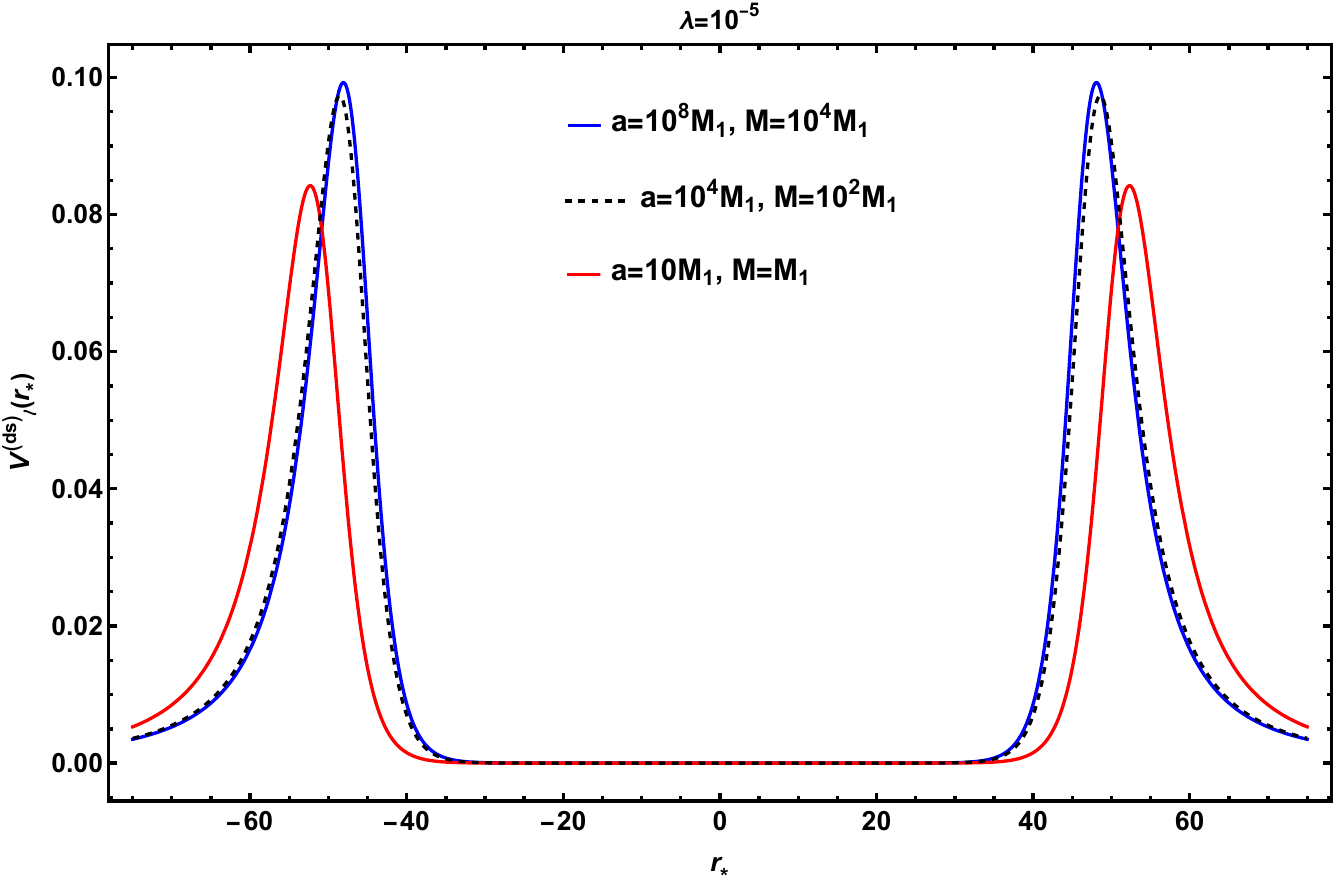}
	\endminipage
	\caption{We have plotted the potential $V_{l}^{\rm (ds)}$ associated with scalar perturbation of the galactic \DS wormhole with the tortoise coordinate $r_{*}$ for different choices of the galactic parameters, with $\ell=1$. As evident, the total potential $V_{l}^{\rm (ds)}$ neatly separates into two individual potentials associated with each of the universes at both sides of the throat. The plot of the potential also depicts that as the ratio $(M/a)$ decreases, both the maxima of the potential increase, while the throat length decreases.}
	\label{fig-Damour-Pot}
\end{figure}

\subsection{Stability of the galactic Damour-Solodukhin wormhole}\label{sec-3.1}

We start by discussing the stability of the galactic \DS wormhole spacetime. In this case, the mass profile $m(r)$ is given by \ref{Damour-Gal-Mass-1} and the $g_{tt}$ component by \ref{Damour-gal-gtt}. Substituting both of these expressions in \ref{gen_pot}, the following explicit form of the potential for the galactic Damour-Solodukhin wormhole can be obtained (the variation of the potential with the radial coordinate has been presented in \ref{fig-Damour-Pot}),
\begin{align}\label{Damour-Potential}
V_{l}^{(\textrm{ds})}(r)=\frac{e^{\gamma}}{r^{2}(a+r)^3}\left[V_{0}(r)+\lambda^{2}V_{\lambda}(r)\right]~,
\end{align}
where $\gamma$ has been defined in \ref{Damour-gal-gtt}, and the expressions for the functions $V_{0}(r)$ and $V_{\lambda}(r)$, respectively, are given by,
\begin{align}\label{damour-v0-and-v-lambda}
V_{0}&=2\left(1-\frac{2M_{1}}{r}\right)\left[M(r-2M_{1})(r-4M_{1})+\left(\frac{M_{1}}{r}+\frac{l(l+1)}{2}\right)(r+a)^{3}-2aMM_{1}\left(1-\frac{2M_{1}}{r}\right) \right]~,
\\
V_{\lambda}&=\frac{M_{1}}{r}\left[\left(1-\frac{4M_{1}}{r}\right)\{(a+r)^{3}-2Ma(r-2M_{1})\}+ 2M (3r-8M_{1})(r-2M_{1})\right]~.
\end{align}
It is possible to provide a simple interpretation for both of these terms $V_{0}$ and $V_{\lambda}$ --- if we had considered a galaxy with a central black hole, then only the term $V_{0}$ would appear, on the other hand, the function $V_{\lambda}$ appears because the central compact object is a wormhole in the present scenario.

\begin{figure}
	\centering
	\minipage{0.5\textwidth}
	\includegraphics[width=\linewidth]{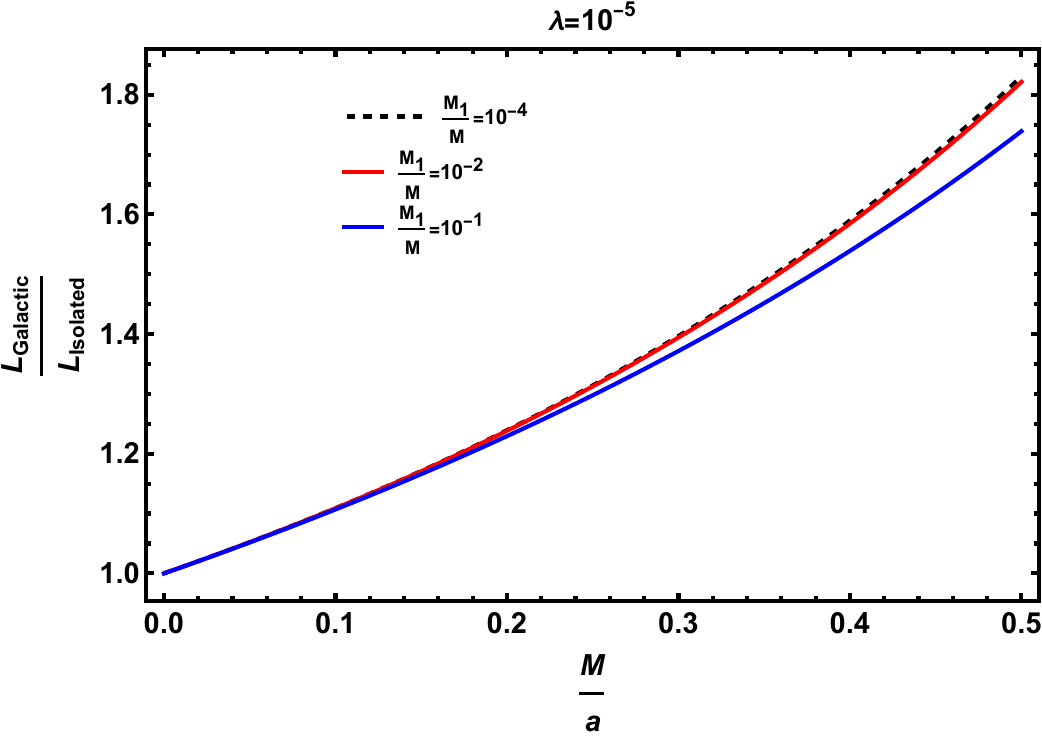}
	\endminipage\hfill
	\caption{This plot shows how the presence of a galaxy influences the throat length. Here we have plotted the ratio of throat length between that of the galactic Damour-Solodukhin wormhole and that of the isolated counterpart as a function of $\frac{M}{a}$; for various values of $\frac{M_{1}}{M}$. From this figure we can see that in the presence of a galaxy, the throat length of a galactic wormhole always increases from that of an isolated counterpart. }\label{fig-Damour-throat-length}
\end{figure}

To proceed further, besides the potential, we need to provide an explicit expression for the tortoise coordinate $r_{*}$ as well. Since the determination of the tortoise coordinate involves an integration, there is a constant of integration present in the analysis. We choose the integration constant, keeping in mind that the wormhole geometry connects two universes by a throat-like structure at $r=2M_{2}$, such that $r_{*}(r=2M_{2})=0$. In this case, we will use $r_{*}\in (-\infty,\infty)$ to cover both the universes, such that $r\rightarrow \pm \infty$ represents the asymptotic region of both the universes. Using the condition that $a>M>>M_{2}>0$ and assuming small values of $\lambda$, we obtain,
\begin{align}\label{appx-tor}
r_{*}\simeq \exp\left[\frac{-\beta}{2}\right] \bigg(1-\frac{2M M_{1}}{a^{2}}\bigg)\left[ r+ 2M_{1}\ln\bigg(\frac{r}{2M_{1}}-1\bigg)\right]
+\frac{L}{2}~.
\end{align}
In the above expression, we have introduced $\beta$ as a shorthand for the quantity, 
\begin{align}\label{def_beta}
\beta=\sqrt{\frac{M}{2a-M}}\left[2\tan^{-1}\left(\frac{a-M}{\sqrt{M(2a-M)}}\right)-\pi\right]~,
\end{align}
and $L$ is referred to as the throat length \cite{Bueno:2017hyj} and physically represents the distance between the two maxima of the perturbing potential, presented in \ref{Damour-Potential}. Therefore, it is a characteristic length scale associated with the scalar perturbation of the galactic \DS wormhole. An explicit expression for the throat length takes the following form, 
\begin{align}\label{Throat}
L=\exp\left[\frac{-\beta}{2}\right]\bigg(1-\frac{2M M_{1}}{a^{2}}\bigg)\left[-4M_{1}+4M_{1}\ln\bigg(\frac{4}{\lambda^{2}}\bigg)\right]~,
\end{align}
where the last part, which is independent of $M$, corresponds to the throat length of the isolated \DS wormhole. Since $\beta$ is always a negative quantity, it follows that the throat length for a galactic \DS wormhole is larger than the throat length of an isolated \DS wormhole (see \ref{fig-Damour-throat-length}). 

\subsubsection{Ringdown spectrum of galactic Damour-Solodukhin wormhole}\label{sec-3.1.2}

Having laid down all the details regarding the perturbation equation associated with the ringdown spectrum, in this section we will discuss the method for obtaining the quasi-normal mode (QNM) frequencies and hence the time domain ringdown signal, which will be compared with the case of an isolated \DS wormhole. For determining the QNM frequencies we can replace the double bump potential barrier (as shown in \ref{fig-Damour-Pot}) by two single bump potentials at both sides of the wormhole throat. Each of these single bump potentials can be thought of as the angular momentum barrier associated with the photon sphere of some black hole spacetime. Mathematically this is achieved by expressing the potential due to scalar perturbation as,
\begin{align}\label{Potential-replacement}
V_{l}^{\textrm{(ds)}}(r_{*})=\theta(r_{*})V_{l}{}_{\rm (single)}\left(r_{*}-\frac{L}{2}\right)+\theta(-r_{*})V_{l}{}_{\rm (single)}\left(-r_{*}-\frac{L}{2}\right)~,
\end{align}
where $V_{l}{}_{\rm (single)}(r_{*})$ is the single bump potential, entering the following differential equation,
\begin{align}\label{Single-BH-master}
\frac{d^{2}\widetilde{\psi}_{lm}}{dr_{*}^{2}}+\Big[\omega^{2}-V_{l}{}_{\rm (single)}(r_{*})\Big]\widetilde{\psi}_{lm}=0~.
\end{align}
Here $\widetilde{\psi}_{lm}(r_{*},\omega)$ is the master variable if we consider only one universe on either side of the wormhole throat located at $r_{*}=0$. From the above structure of the potential $V_{l}$, it follows that $V_{l}{}_{\rm (single)}(r_{*})$ vanishes at both the asymptotic and in the near throat region, yielding the following nature for the perturbation $\widetilde{\psi}_{lm}(r_{*},\omega)$,
\begin{align}\label{psi-u-nature} 
\widetilde{\psi}_{lm}(r_{*},\omega)=
\begin{cases} 
Ae^{-i\omega r_{*}}+Be^{i\omega r_{*}} & \text{for}\ \quad r_{*}\to \infty
\\
Ce^{-i\omega r_{*}}+De^{i\omega r_{*}} & \text{for} \quad r_{*}\to 0~.
\end{cases} 
\end{align}
Note that here $r_{*}\rightarrow 0$ does not correspond to the horizon, but rather the near-throat region, so that both the ingoing and the outgoing modes will exist. Further, it follows that $V_{l}{}_{\rm (single)}(r_{*})$ becomes $\mathcal{O}(\lambda^{2})$ at the location of the throat, i.e., at $r=2M_{2}$, which is several orders of magnitude smaller than the frequency $\omega$ and hence we can approximate $\sqrt{\omega^{2}-V_{l}{}_{\rm (single)}(0)}\approx \omega$. At this point, we introduce the $(2\times 2)$ transfer matrix $T$ such that it relates the near-throat and asymptotic amplitudes of the incoming and outgoing waves in the region $r_{*}>0$, yielding,
\begin{align}\label{Single-transfer-matrix}
\begin{pmatrix}
B\\
A
\end{pmatrix}=T\begin{pmatrix}
D\\
C
\end{pmatrix}~.
\end{align}
The above coefficients appearing with the ingoing and the outgoing wave modes can be further constrained by imposing appropriate boundary conditions consistent with the QNMs, which correspond to --- there are no incoming waves from the asymptotic regions of both universes. For the universe with $r_{*}>0$, this condition is manifested by setting $A=0$ in \ref{Single-transfer-matrix}, which leads to the condition,
\begin{align}\label{BH-Reflectivity}
\frac{C}{D}\equiv R_{\rm (single)}(\omega)=-\frac{T_{21}}{T_{22}}~,
\end{align}
Here $R_{\rm (single)}(\omega)$ is the reflectivity of the single bump potential when probed from the throat towards the asymptotic region. Similarly, one can study the scattering problem associated with the universe having $r_{*}<0$. In this context as well, there exists a single bump potential on the negative $r_{*}$ axis (see \ref{fig-Damour-Pot}), for which the notion of the ingoing and the outing modes reverses with respect to that of $r_{*}>0$. Keeping this in mind, one can find the transfer matrix $\widetilde{T}$ for the universe with $r_{*}<0$ by using the transfer matrix $T$ associated with the universe having $r_{*}>0$, which is given by,
\begin{align}\label{Transfer-Other}
\widetilde{T}=\sigma_{x}T^{-1}\sigma_{x}~.
\end{align}
We now have both the transfer matrices associated with the two universes separately, as a final step we need to match the near-throat amplitudes of both universes to obtain the full transfer matrix ($\mathbb{T}$) for the scattering problem associated with the potential $V_{l}^{\textrm{(ds)}}(r_{*})$. This yields,
\begin{align}\label{tranfer-V-WH}
\mathbb{T}=T\begin{pmatrix}
e^{i\omega L}&0\\
0&e^{-i\omega L}
\end{pmatrix}\widetilde{T}~,
\end{align} 
where the matrix in the middle is a simple translation matrix, relating the amplitudes of outgoing and ingoing modes beyond the single bump potential in one universe to that in another. As already emphasized above, the role of this transfer matrix is to relate the asymptotic amplitudes of the universe with $r_{*}>0$ to that of asymptotic amplitudes of the universe with $r_{*}<0$. 

\begin{table}[th!]
	\centering
	\def\arraystretch{1.3}
	\setlength{\tabcolsep}{1.5em}
	\begin{tabular}{|p{1cm}||p{5cm}|p{5cm}|  }
	\hline
\multicolumn{3}{|c|}{Comparison of QNM frequencies} \\
	\hline
	Mode $ n $    & For isolated wormhole      & For galactic wormhole     \\ \hline
		$ 1  $    & $0.0321-1.083.10^{-4}i $  & $0.0326-1.2174.10^{-4}i$   \\ \hline
		$ 2   $   & $0.0638-5.720.10^{-4}i$   & $0.0649-7.6526.10^{-4}i$  \\ \hline
		$ 3     $ & $0.0950-1.679.10^{-3}i $  & $0.0967-1.9841.10^{-3}i$ \\ \hline
		$ 4    $  & $0.1258-3.709.10^{-3}i$   & $0.1280-4.0441.10^{-3}i$  \\ \hline
		$ 5     $ & $0.1566-6.739.10^{-3}i $  & $0.1592-7.0129.10^{-3}i$ \\ \hline
		$ 6    $  & $0.1877-1.0106.10^{-2}i $  & $0.1906-1.0746.10^{-2}i$ \\ \hline
    	$ 7     $ & $0.2192-1.490.10^{-2}i$   & $0.2222-1.4990.10^{-2}i$  \\ \hline
		$ 8   $   & $0.2510-1.948.10^{-2}i $  & $0.2541-1.9518.10^{-2}i$   \\ \hline
		$ 9    $  & $0.2832-2.418.10^{-2}i $  & $0.2862-2.4185.10^{-2}i$   \\ \hline
		$ 10  $   & $0.3158-2.90.10^{-2}i $  & $0.3185-2.8916.10^{-2}i$   \\ \hline
	\end{tabular}
	\caption{A Comparison of the dimensionless QNM frequencies between the isolated and the galactic Damour-Solodukhin wormhole has been presented for $l=0$ and $\lambda=10^{-5}$. The galactic parameters are chosen as: $a=10^{8}M_{1}$, and $M=10^{4}M_{1}$. As evident from the table, within the same precision, the real part of the QNM frequencies of both wormholes are in agreement, but the magnitude of the imaginary part of the QNM frequencies for the galactic \DS wormhole is slightly larger than that of the isolated one. Suggesting that the galactic \DS wormhole is more stable than the isolated one.}\label{Damour-QNM-COMP-l0}
\end{table}

\begin{table}[th!]
	\centering
	\def\arraystretch{1.3}
	\setlength{\tabcolsep}{1.5em}
	\begin{tabular}{|p{1cm}||p{5cm}|p{5cm}|  }
	\hline
\multicolumn{3}{|c|}{Comparison of QNM frequencies} \\
	\hline
	Mode $ n $    & For isolated wormhole      & For galactic wormhole     \\ \hline
		$ 1  $    & $0.0372-7.9961.10^{-7}i $  & $0.0372-8.0010.10^{-7}i$   \\ \hline
		$ 2   $   & $0.0741-4.2698.10^{-6}i$   & $0.0741-4.2730.10^{-6}i$  \\ \hline
		$ 3     $ & $0.1102-1.4880.10^{-5}i $  & $0.1102-1.4893.10^{-5}i$ \\ \hline
		$ 4    $  & $0.1455-4.5449.10^{-5}i$   & $0.1455-4.5498.10^{-5}i$  \\ \hline
		$ 5     $ & $0.1798-1.2987.10^{-4}i $  & $0.1798-1.3002.10^{-4}i$ \\ \hline
		$ 6    $  & $0.2132-3.5057.10^{-4}i $  & $0.2132-3.5103.10^{-4}i$ \\ \hline
    	$ 7     $ & $0.2455-8.8235.10^{-4}i$   & $0.2455-8.8357.10^{-4}i$  \\ \hline
		$ 8   $   & $0.2767-2.0175.10^{-3}i $  & $0.2768-2.0202.10^{-3}i$   \\ \hline
		$ 9    $  & $0.3072-4.0814.10^{-3}i $  & $0.3073-4.0865.10^{-3}i$   \\ \hline
		$ 10  $   & $0.3375-7.2239.10^{-3}i $  & $0.3375-7.2318.10^{-3}i$   \\ \hline
	\end{tabular}
	\caption{A comparison between the real and the imaginary parts of the QNM frequencies of the isolated and the galactic \DS wormhole has been presented for $\ell=1$ and $\lambda=10^{-5}$. The galactic parameters are chosen as $a=10^{8} M_{1}$, and $M=10^{4}M_{1}$. In this case also the values are more or less in agreement, with the galactic wormhole being more stable than the isolated one.}\label{Damour-QNM-COMP-l1}
\end{table}

\begin{figure}
	\centering
	\minipage{0.5\textwidth}
	\includegraphics[width=\linewidth]{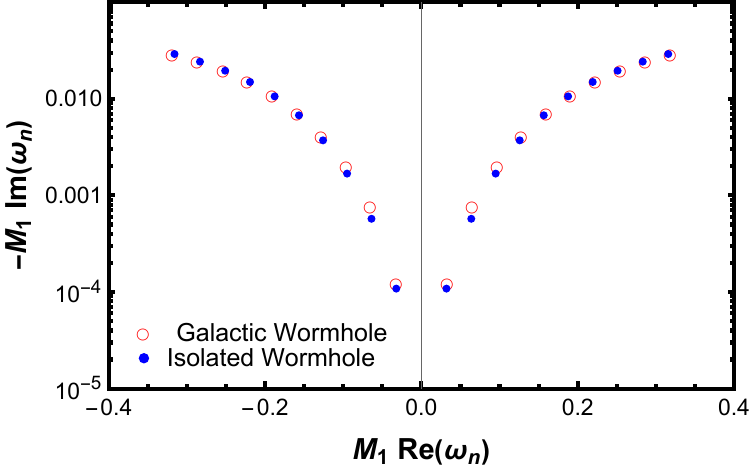}
	\endminipage\hfill
	\minipage{0.5\textwidth}
	\includegraphics[width=\linewidth]{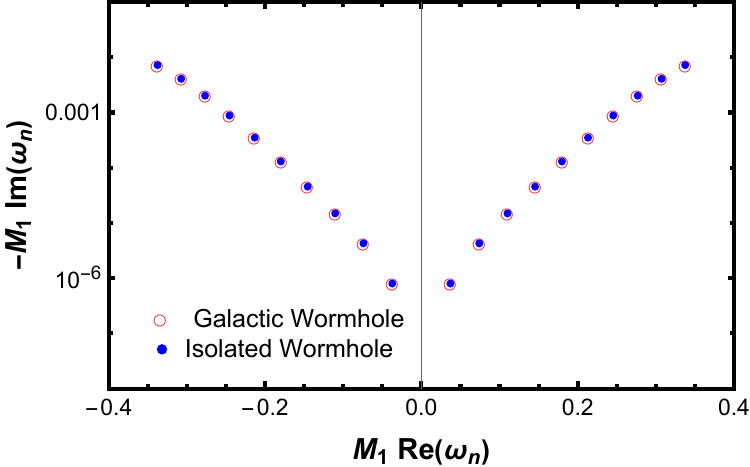}
	\endminipage
	\caption{The real and the imaginary parts of the QNM frequencies of the galactic \DS wormhole has been depicted in the left figure for $\ell=0$, with the metric parameters chosen to be: $a=10^{8}M_{1}$, $M=10^{4}M_{1}$, and $\lambda=10^{-5}$. While the plot to the right describes the real and the imaginary parts of the QNM frequencies for $\ell=1$ with identical choices of the metric parameters.}\label{fig-Damour-QNM}
\end{figure}	

Since we have already imposed the condition that there are no incoming waves from $r_{*}=+\infty$, it is time to impose the other condition that there are no incoming waves from $r_{*}=-\infty$ as well, which demands $\mathbb{T}_{22}=0$, we obtain the following condition from \ref{tranfer-V-WH},
\begin{align}\label{Wormhole-QNF}
e^{-i\omega_{n}L}=-e^{-in\pi}R_{\rm (single)}(\omega)~,n=1, 2,~...~,
\end{align}
where $L$ is the throat length and $R_{\rm (single)}(\omega)$ is the reflectivity of a single bump potential. The solution of the above equation yields the associated QNM frequencies. For determining the QNM frequencies we first find out the reflectivity of the single bump potential numerically and then use \ref{Wormhole-QNF}. The result of this analysis has been presented in \ref{fig-Damour-QNM} as well as in \ref{Damour-QNM-COMP-l0} and \ref{Damour-QNM-COMP-l1}. In particular, \ref{fig-Damour-QNM} depicts explicitly the real and imaginary parts of the QNM frequencies for the galactic \DS wormhole spacetime, which we have compared with that of an isolated \DS wormhole. We observe that the order of magnitudes of the real and the imaginary parts of the QNM frequencies of the galactic \DS wormhole are in consonance with that of the isolated \DS wormhole. However, the imaginary parts of the QNM frequencies for galactic \DS wormholes are larger than their isolated counterpart, implying that in the presence of a galaxy, the wormhole is more stable than its isolated counterpart. This is also evident from both \ref{Damour-QNM-COMP-l0} and \ref{Damour-QNM-COMP-l1} for different choices of the angular number $l$. 

To obtain the time-domain signal for the ringdown profile of the galactic \DS wormhole, we need to analyze the evolution of the scalar perturbation in the time domain. For this purpose, we need to define the Fourier counterpart of $\psi_{lm}(r_{*},\omega)$ in the usual manner. Using the perturbation in the time domain and the fact that for a static and spherically symmetric spacetime, the effective potential appearing in \ref{Damour-Radial-Master} is frequency independent, we can rewrite the radial perturbation equation, presented in \ref{Damour-Radial-Master}, in the time domain as follows,
\begin{align}\label{Time-domain-EQ}
-\frac{\partial^2\psi_{lm}(r_{*},t)}{\partial t^2}+\frac{\partial^{2}\psi_{lm}(r_{*},t)}{\partial r_{*}^{2}}-V_{l}^{\textrm{(ds)}}(r_{*})\psi_{lm}(r_{*},t)=0~.
\end{align}

\begin{figure}
	\centering
	\minipage{0.8\textwidth}
	\includegraphics[width=\linewidth]{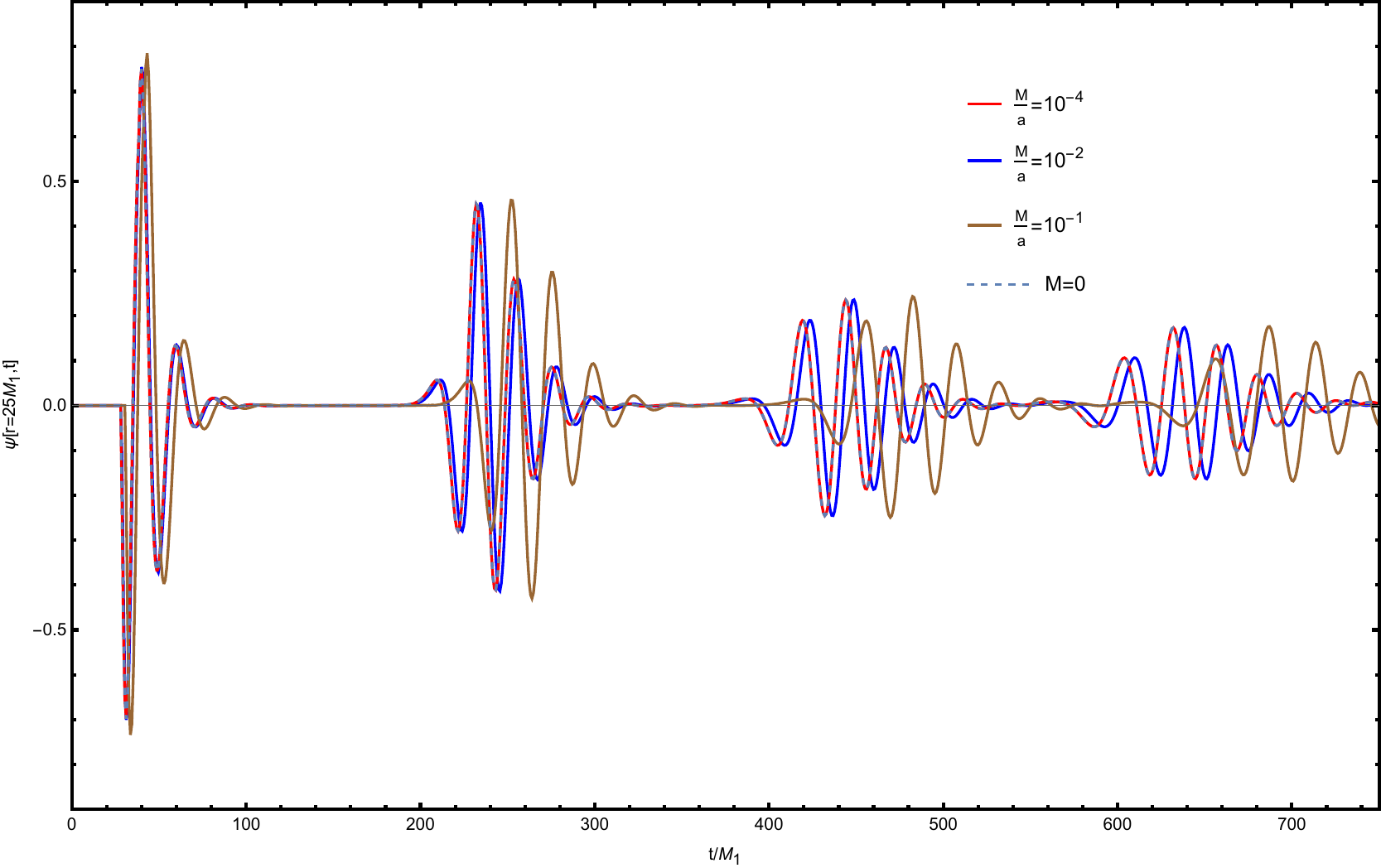}
	\endminipage\hfill
	\caption{The time domain ringdown waveform for scalar perturbation has been presented for galactic and isolated \DS wormhole at $r=25 M_{1}$ for $\ell=1$ with $\lambda=10^{-5}$. We have chosen various values of $\frac{M}{a}$ for the galactic wormhole. In this case, $M=0$ corresponds to the isolated wormhole. From the plot, one can see that except for $\frac{M}{a}=10^{-1}$, the primary signal of the waveform matches very well with the isolated wormhole. Moreover, the late time echoes give us the scope to determine the galactic parameters via the echo time delay. The parameters appearing in the initial Gaussian profile (see \ref{Initial conditions}) are chosen such that $r_{*}^{0}$ is larger than the location of the photon sphere and $\sigma=10M_{1}$. Using this choice, we have numerically solved the time domain wave equation and presented it here.}\label{fig-Damour-echo}
\end{figure}

\begin{figure}
	\centering
	\minipage{0.8\textwidth}
	\includegraphics[width=\linewidth]{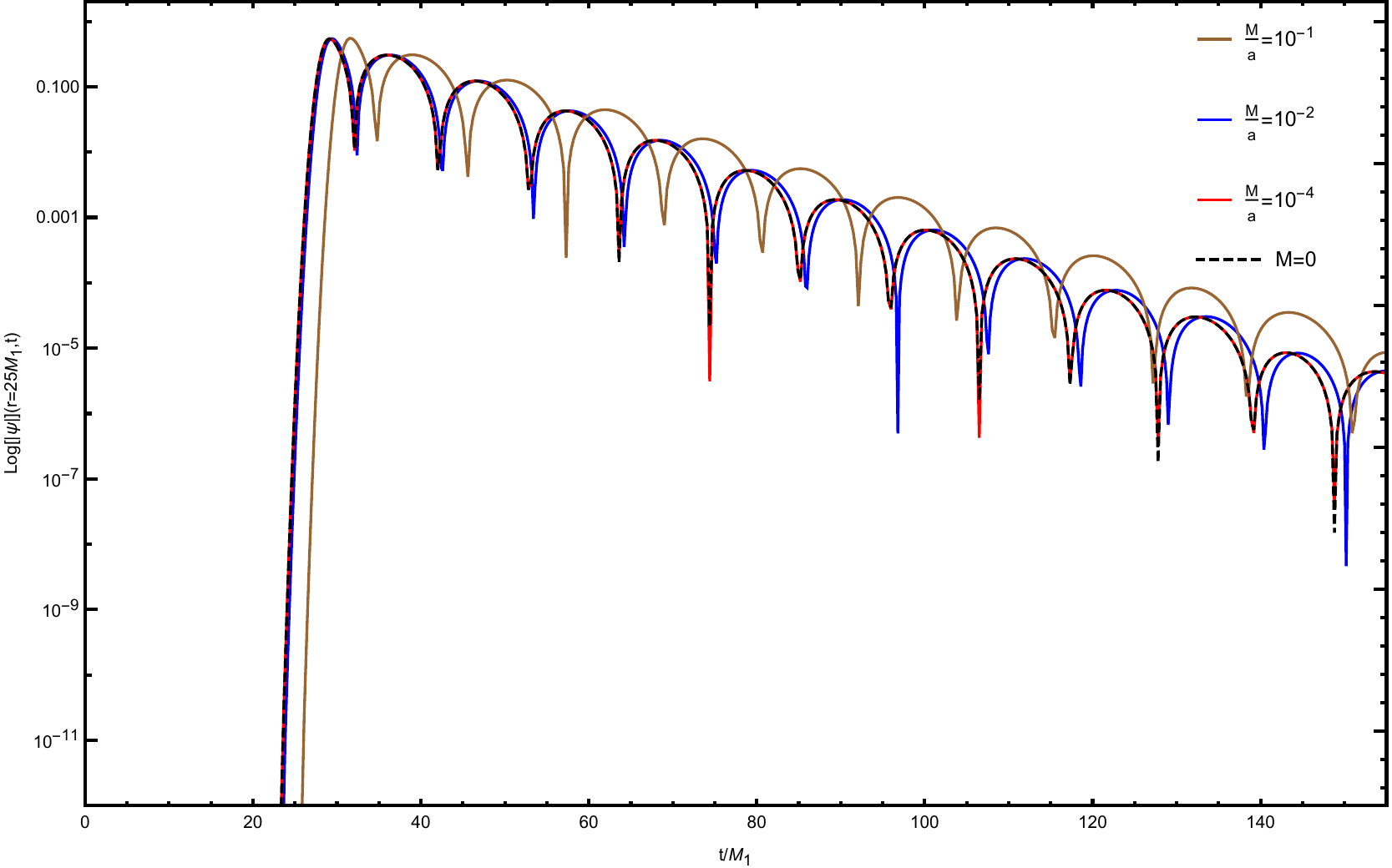}
	\endminipage\hfill
	\caption{To distinguish the prompt ringdown signal in the waveform of galactic \DS wormhole from that of an isolated wormhole, we have plotted the logarithm of the absolute value of the ringdown signal for $l=1$ with $\lambda=10^{-5}$ at $r=25M_{1}$.  The parameters of the plot are chosen such that $\sigma=M_{1}$ and $r^{0}_{*}$ is larger than the photon sphere. From the plot, one can see that except for large $(M/a)$ ratio for the galactic matter, e.g., $\frac{M}{a}=10^{-1}$, the primary signal of the waveform matches very well with that of the isolated wormhole.}\label{fig-Damour-Log}
\end{figure}

This is a second-order partial differential equation in $r_{*}$ and $t$, and hence to solve this equation, one must specify two initial conditions in time and two boundary conditions in the tortoise coordinate. The outgoing boundary conditions at two asymptotic infinities can be implemented as  $\partial_{r_*}\psi_{lm}(r_{*},t)= \mp \partial_{t}\psi_{lm}(r_{*},t)$, as $r_*\to \pm \infty$. While we choose the following initial conditions:
\begin{equation}\label{Initial conditions}
\psi_{lm}(r_*,0)=0\quad\text{and}\quad\partial_{t}\psi_{lm}(r_{*},0)=e^{-\frac{(r_{*}-r^{0}_{*})^{2}}{\sigma^{2}}}~,
\end{equation}
where $r^{0}_{*}$ and $\sigma^{2}$ will be chosen in such a way that the primary signal appears outside of the unstable photon sphere, i.e., the maxima of the single bump potential. Using these boundary and initial conditions, we solve \ref{Time-domain-EQ} numerically. The result of this analysis can be found from both \ref{fig-Damour-echo} and \ref{fig-Damour-Log}, where we have shown the ringdown waveform in the time domain, which consists of echos of the primary signal due to repetitive reflection of primary signal between the two maxima of the potential $V_{l}^{\rm (ds)}(r_{*})$. The time delay between two successive echos is given by $\Delta t=2L$. As in the presence of galactic matter, the throat length increases, as a consequence the time delay also increases. This can be seen explicitly from \ref{Throat}. Therefore, if echoes are observed in the time domain waveform, they can provide crucial hints not only for the central wormhole but also about the dark matter halo surrounding it.   

\subsection{Ringdown profile and stability of galactic braneworld wormhole}\label{sec-3.2}

\begin{figure}
	\centering
	\minipage{0.7\textwidth}
	\includegraphics[width=\linewidth]{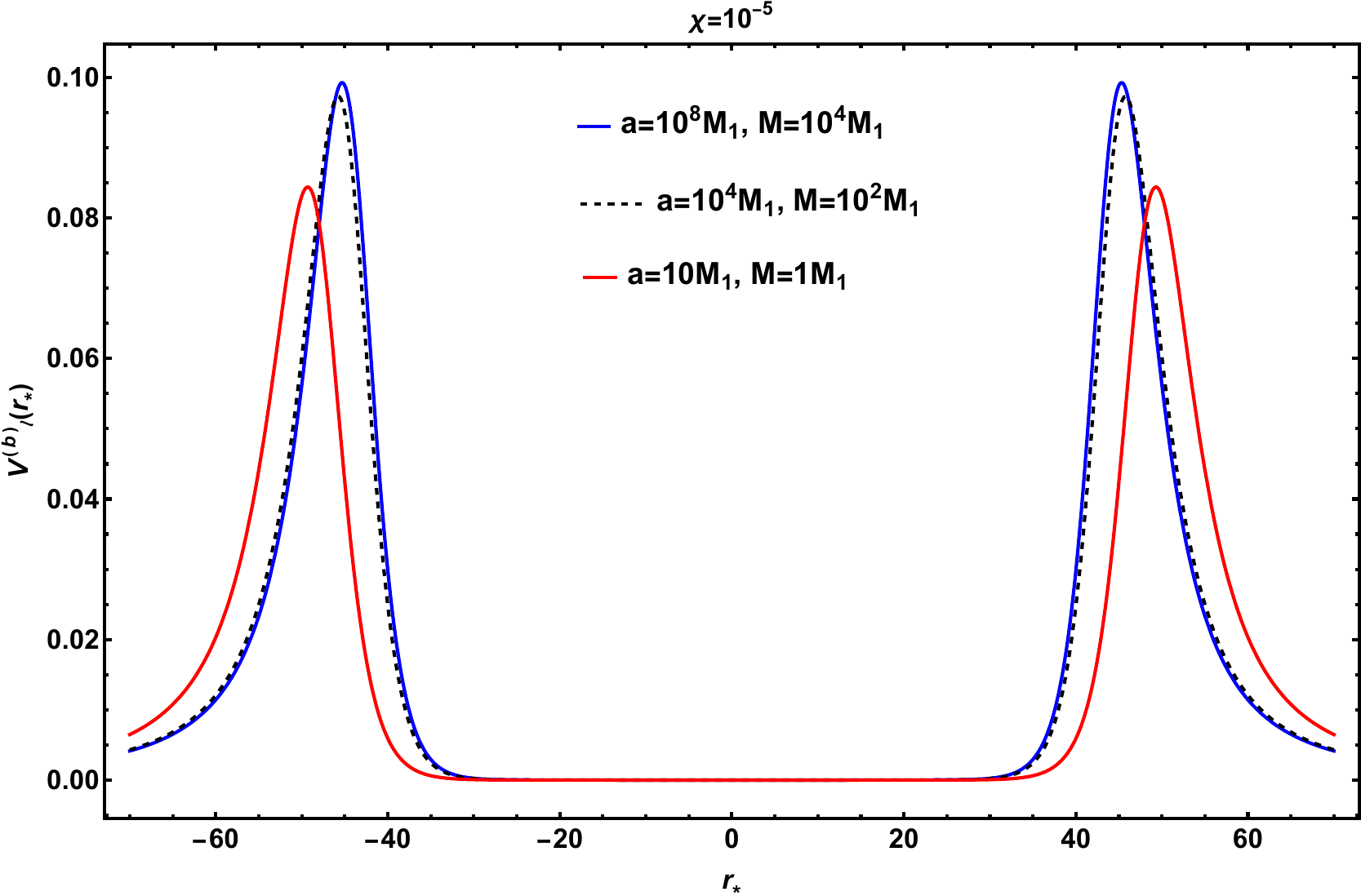}
	\endminipage
	\caption{We have plotted the potential $V_{l}^{(\textrm{b})}$ associated with scalar perturbation of the galactic braneworld wormhole with the tortoise coordinate $r_{*}$ for different choices of the galactic parameters, with $\ell=1$. As evident, the total potential $V_{l}^{(\textrm{b})}$ neatly separates into two individual potentials associated with each of the universes at both sides of the throat. The plot of the potential also depicts that as the ratio $(M/a)$ decreases, both the maxima of the potential increases and the distance between the two maxima decreases.}
	\label{fig-Brane-Pot}
\end{figure}

\begin{figure}
	\centering
	\minipage{0.5\textwidth}
	\includegraphics[width=\linewidth]{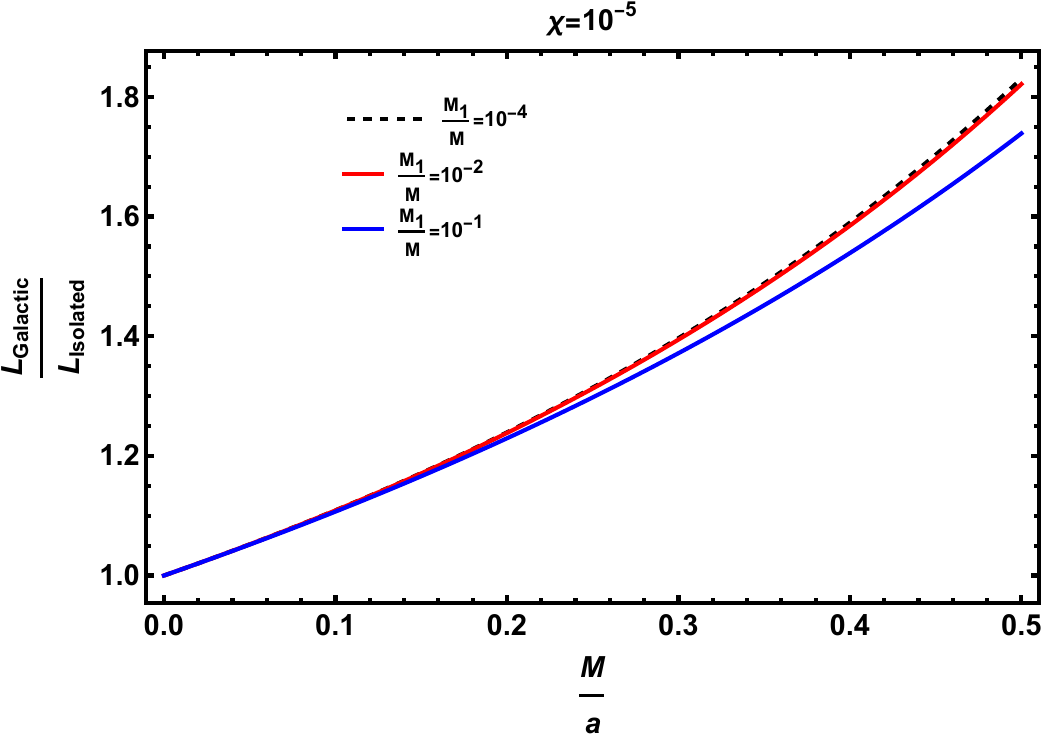}
	\endminipage\hfill
	\caption{This plot shows how the presence of a galaxy influences the throat length of the braneworld wormhole. Here we have plotted the ratio of throat length between that of the braneworld galactic wormhole and that of the isolated counterpart as a function of $\frac{M}{a}$; for various values of $\frac{M_{1}}{M}$. From this figure, we can see that in the presence of a galaxy, the throat length of a galactic wormhole always increases from that of an isolated counterpart, as in the case of \DS wormhole.}\label{fig-Brane-throat-length}
\end{figure}

In this section, we will present the QNM frequencies and the ringdown profiles in the time domain using the method introduced in \ref{sec-3.1.2}, but for the galactic braneworld wormhole. In this case, also the evolution of the perturbing scalar field $\Psi$ is described by the massless Klein-Gordon equation, but we have to make an additional assumption that the energy density of the field $\Psi$ is much smaller than the energy density of the background radion incarnation $\Phi$, loosely speaking $|\Psi|<<|\Phi|$ (see \cite{PhysRevD.106.124003} for further details). Like the previous case, here also the potential experienced by the scalar perturbation $\Psi$ can be expressed in terms of galactic as well as wormhole parameters. However, the potential has a very complicated structure, and hence we have delegated its analytic expression to \ref{app-D}. But, the structure of the potential has been plotted in \ref{fig-Brane-Pot} for different choices of the galactic parameters, and as evident from the plots, as the dimensionless ratio $(M/a)$ decreases the distance between the maxima decreases while the maxima themselves increase. 
 The tortoise coordinate used in the master equation will be chosen in such a way that $r_{*}(r=2M_{1})=0$, and we will have $r_{*}\in (-\infty,+\infty)$ to cover both the asymptotically flat spacetimes on each side of the throat. As in the previous scenario, for the case of the braneworld wormhole as well we can not provide an explicit analytic expression for the tortoise coordinate in terms of the radial coordinate $r$, but assuming $a>M>>M_{1}>0$ and $\chi\gtrsim 0$, we obtain,
\begin{align}\label{brane-appx-tor}
r_{*}\simeq \exp\left[\frac{-\beta}{2}\right] \bigg(1-\frac{2M M_{1}}{a^{2}}\bigg)\left[ r+ 2M_{1}\ln\bigg(\frac{r}{2M_{1}}-1\bigg)\right]+\frac{\widetilde{L}}{2}~.
\end{align}
The quantity $\beta$ has already been defined in \ref{def_beta} and the throat length $\widetilde{L}$ is given by,
\begin{align}\label{brane-throat-length}
\tilde{L}=-4M_{1}\exp\left[\frac{-\beta}{2}\right]\bigg(1-\frac{2M M_{1}}{a^{2}}\bigg)\left(1+2\ln \chi\right)~.
\end{align}
Since the wormhole parameter $\chi$ is taken to be smaller than unity, the throat length is actually positive. Also, similar to the case of a galactic \DS wormhole, in the case of a galactic braneworld wormhole the throat length increases in the presence of a galactic halo, see \ref{fig-Brane-throat-length}. 

\begin{table}[th!]
	\centering
	\def\arraystretch{1.3}
	\setlength{\tabcolsep}{1.5em}
	\begin{tabular}{|p{1cm}||p{5cm}|p{5cm}|  }
	\hline
\multicolumn{3}{|c|}{Comparison of the QNM frequencies} \\
	\hline
	Mode $ n $    & For isolated wormhole      & For galactic wormhole     \\ \hline
		$ 1  $    & $0.0398-9.9449.10^{-7}i $  & $0.0394-9.8454.10^{-7}i$   \\ \hline
		$ 2   $   & $0.0792-5.5084.10^{-6}i$   & $0.0784-5.4532.10^{-6}i$  \\ \hline
		$ 3     $ & $0.1176-2.0163.10^{-5}i $  & $0.1164-1.9961.10^{-5}i$ \\ \hline
		$ 4    $  & $0.155-6.4867.10^{-5}i$   & $0.1534-6.4218.10^{-5}i$  \\ \hline
		$ 5     $ & $0.1912-1.9426.10^{-4}i $  & $0.1893-1.9232.10^{-4}i$ \\ \hline
		$ 6    $  & $0.2262-5.4270.10^{-4}i $  & $0.2239-5.3728.10^{-4}i$ \\ \hline
    	$ 7     $ & $0.2599-1.3837.10^{-3}i$   & $0.25733-1.3699.10^{-3}i$  \\ \hline
		$ 8   $   & $0.2926-3.1189.10^{-3}i $  & $0.2896-3.0878.10^{-3}i$   \\ \hline
		$ 9    $  & $0.3246-6.0766.10^{-3}i $  & $0.3213-6.0157.10^{-3}i$   \\ \hline
		$ 10  $   & $0.3565-1.0247.10^{-2}i $  & $0.3529-1.0145.10^{-2}i$   \\ \hline
	\end{tabular}
	\caption{We have presented a comparison of the QNM frequencies between the isolated and the galactic braneworld wormhole for $\ell=1$ and $\chi=10^{-5}$. The galactic parameters are chosen as $a=10^{4} M_{1}$, and $M=10^{2} M_{1}$. From the above table it is clear that within the same precision, the magnitude of the imaginary part of the QNM frequencies of the isolated wormhole is slightly larger than that of the galactic one. Therefore the isolated wormhole is more stable than the galactic one.}\label{Brane-QNM-COMP-l1}
\end{table}

\begin{figure}
	\centering
	\minipage{0.5\textwidth}
	\includegraphics[width=\linewidth]{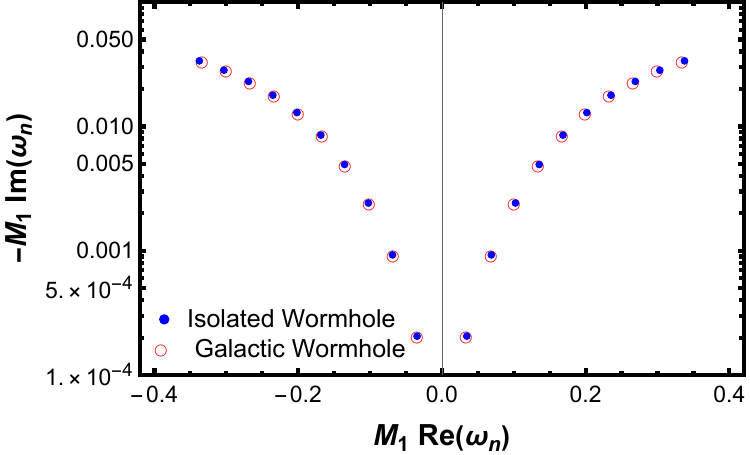}
	\endminipage\hfill
	\minipage{0.5\textwidth}
	\includegraphics[width=\linewidth]{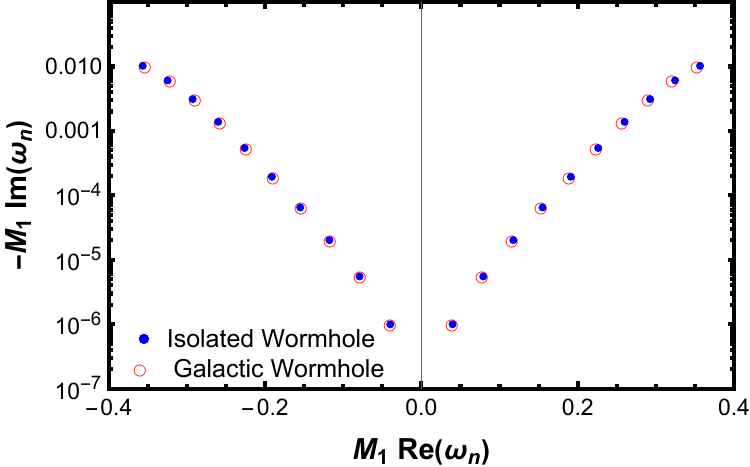}
	\endminipage
	\caption{On the left we have plotted the QNM frequencies for $\ell=0$ mode of the scalar perturbation on the galactic and isolated braneworld wormhole with metric parameters chosen to be: $a=10^{4}M_{1}$, $M=10^{2}M_{1}$, and $\chi=10^{-5}$. While the right-hand side plot of the QNM frequencies is for the same systems, but with $\ell=1$ with metric parameters chosen to be: $a=10^{4}M_{1}$, $M=10^{2}M_{1}$, and $\chi=10^{-5}$.}\label{fig-Brane-QNM}
\end{figure}

  \begin{figure}
	\centering
	\minipage{0.8\textwidth}
	\includegraphics[width=\linewidth]{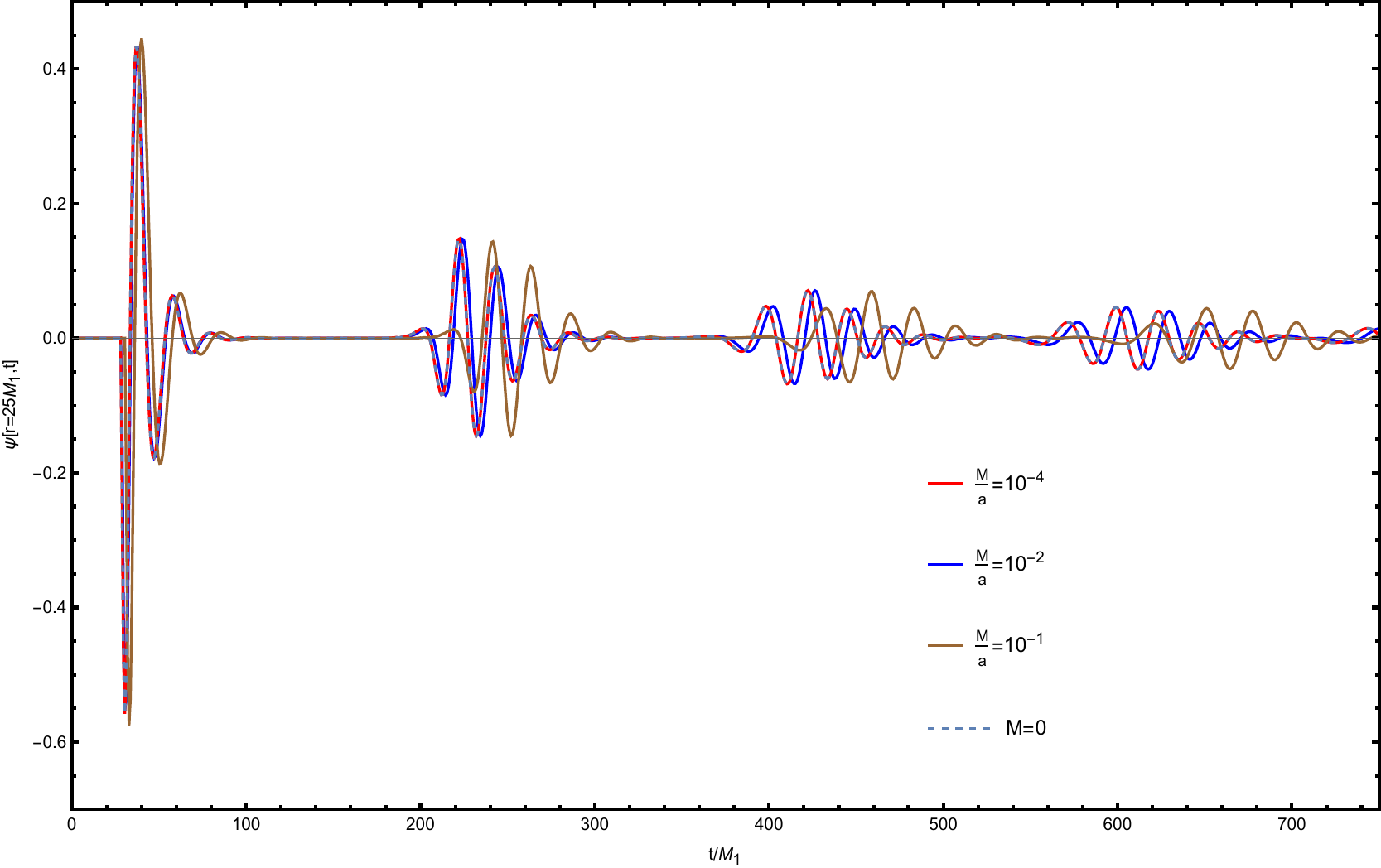}
	\endminipage\hfill
	\caption{The time domain ringdown waveform for scalar perturbation has been presented for galactic and isolated braneworld wormhole at $r=25 M_{1}$ for $\ell=1$ with $\chi=10^{-5}$. For the galactic wormhole, we have chosen various values of $\frac{M}{a}$. In this case, $M=0$ corresponds to the isolated wormhole. From the plot, one can see that except for $\frac{M}{a}=10^{-1}$, the primary signal of the waveform matches very well with the isolated wormhole. Moreover, the late time echoes give us the scope to determine the galactic parameters via the echo time delay. The parameters appearing in the initial Gaussian profile (see \ref{Initial conditions}) are chosen such that $r_{*}^{0}$ is larger than the location of the photon sphere, and $\sigma=10M_{1}$. Using this choice, we have numerically solved the time domain wave equation and presented it here.}\label{fig-Brane-echo}
\end{figure}

  \begin{figure}
	\centering
	\minipage{0.8\textwidth}
	\includegraphics[width=\linewidth]{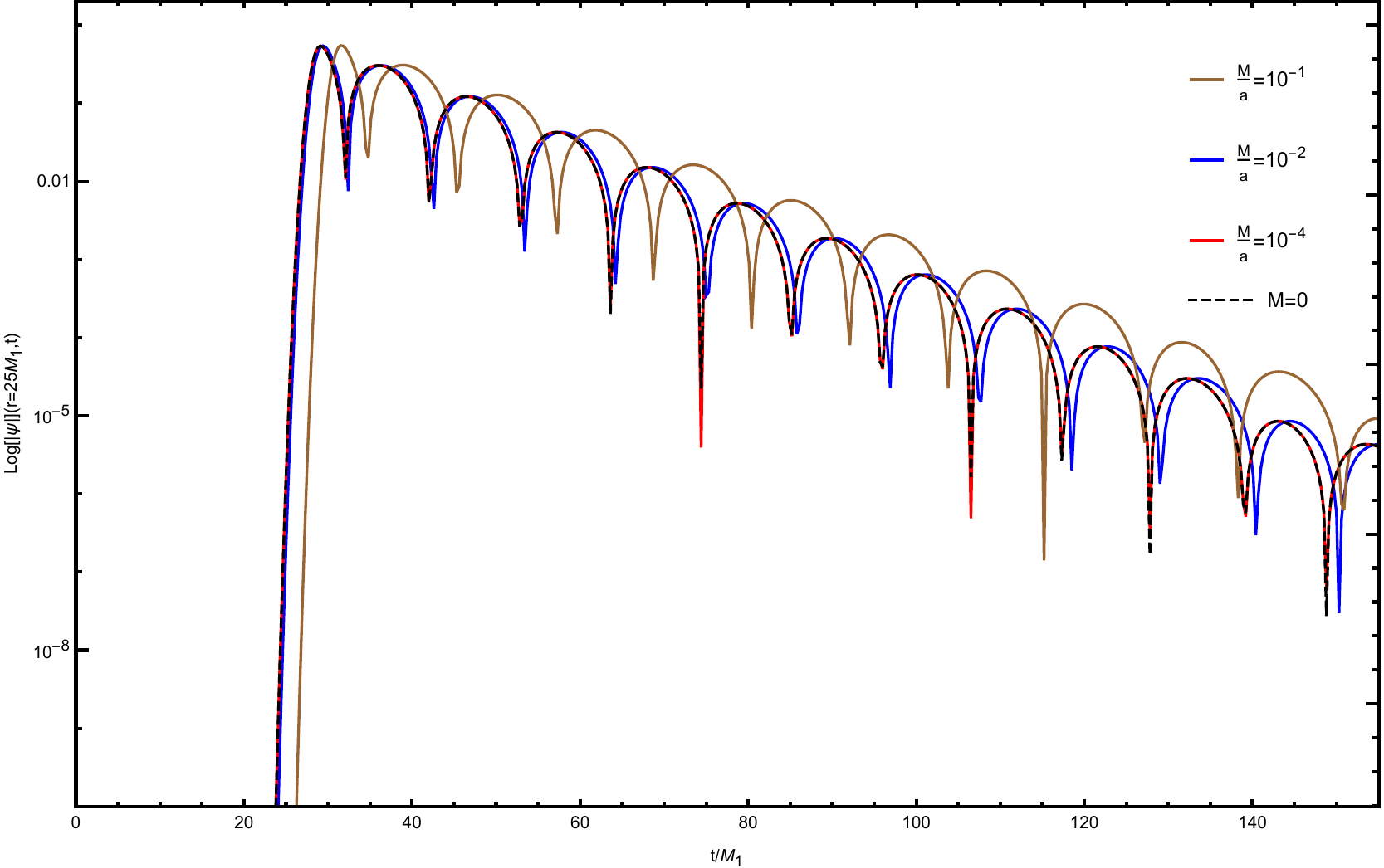}
	\endminipage\hfill
	\caption{To distinguish the primary signal in the ringdown waveform of a galactic braneworld wormhole from that of an isolated wormhole, we have plotted the logarithm of the absolute value of the ringdown signal for $l=1$ with $\chi=10^{-5}$ at $r=25M_{1}$.  The parameters of the plot are chosen such that $\sigma=1M_{1}$ and $r^{0}_{*}$ is larger than the photon sphere.  From the plot, one can see that except for $\frac{M}{a}=10^{-1}$, the primary signal of the waveform matches very well with the isolated wormhole.  }\label{fig-Brane-Log}
\end{figure}

Having outlined the basic setup for the scalar perturbation, one can express the effective potential $V_{l}^{(\textrm{b})}$ for the galactic braneworld wormhole in terms of two single bump potentials appearing on both sides of the wormhole throat. Thus one can employ the same techniques as in the previous section and shall finally obtain the QNM frequencies by solving \ref{Wormhole-QNF} with the reflectivity due to the appropriate potential for the galactic braneworld wormhole. Here also we numerically solve for the reflectivity and hence obtain the QNM frequencies, which have been presented in \ref{fig-Brane-QNM} and \ref{Brane-QNM-COMP-l1}. In \ref{fig-Brane-QNM}, besides presenting the real and imaginary parts of the QNM frequencies associated with the scalar perturbation of the galactic braneworld wormhole, we have also presented the real and imaginary parts of the QNM frequencies of the isolated braneworld wormhole. As evident, the QNM frequencies associated with scalar perturbation of both of these spacetimes have the same order, but their imaginary parts differ in magnitude. In particular, the imaginary part of the QNM frequency of a braneworld wormhole in a galaxy will be smaller than the isolated one, and hence it follows that the isolated braneworld wormhole is more stable compared to the galactic wormhole. This is in contrast to the \DS wormhole. Finally, the determination of the ringdown waveform in the time domain can be obtained by using the method outlined in \ref{sec-3.1.2} with appropriate boundary and initial conditions. The result of such an analysis has been presented in \ref{fig-Brane-echo} and \ref{fig-Brane-Log}. As evident for these plots, the echoes are sensitive to even small values of the ratio  $(M/a)$, however, the prompt ringdown only gets affected for $(M/a)\sim 10^{-1}$.  

\section{Discussion}\label{sec-4}

Wormholes are exotic objects, which can be as compact as that of black holes, but without any event horizon. Since wormhole geometries involve photon sphere, the prompt ringdown of the GW waveform can be mimicked by them, but modifications will come in at later times, with the introduction of echoes in the time domain signal. There has been extensive discussion involving such ECOs, regarding their origin, stability and response under various external perturbations of scalar, electromagnetic and gravitational origin. However, all such discussions were in the context of isolated systems. In this work we have made the first attempt to incorporate environmental effects, mainly that of the surrounding dark matter halo, on the geometry of certain wormhole spacetimes. We hope our strategy will be useful for embedding other ECOs in the surrounding dark matter halo, as well.  

We have used two distinct wormhole solutions for our purpose --- (a) the \DS wormhole and (b) the braneworld wormhole. In both cases, we could provide a suitable mass profile, which predicts wormhole matter distribution at small radii and a Hernquist-type galactic mass profile at a larger radius. Astonishingly, the galactic wormholes, be it \DS or braneworld, have their throat at exactly the same location as that of their isolated counterpart. Moreover, the presence of galactic matter tames the violation of the energy conditions significantly. For example, in the context of an isolated \DS wormhole the null and the strong energy conditions are violated everywhere, while, for the galactic \DS wormhole, the violations of the energy conditions happen in a very small region close to the throat of the wormhole. This result can also be understood in an intuitive manner --- near the throat the properties of the wormhole dominate and hence the energy conditions are violated, while away from the throat, the properties of the galactic matter take over, restoring the energy conditions.  

The impact of galactic matter can also be felt on the physical properties of the wormholes, e.g., the location of the photon sphere, ISCO, and shadow radius. In particular, it generically follows that the shadow radius and the photon sphere radius for galactic wormholes are larger than their isolated counterpart. This means galactic wormholes cast a larger shadow. In this respect, it is worth pointing out that the shadow radius of the M87$^{*}$ within the M87 galaxy is larger than what is expected from the dynamics of stars and gases surrounding it. A possible reason for the larger shadow could be due to the presence of a dark matter halo surrounding the same, given our findings regarding the shadow radius in this work\footnote{Such an enhancement in the shadow radius can also be due to other effects, e.g., the presence of extra dimensions can also enhance the shadow radius through a negative tidal charge parameter \cite{Banerjee:2022jog, Banerjee:2019nnj}.}. Similarly, the enhancement of the ISCO radius will also have implications for accretion physics. Since the luminosity arising from the accretion disc is obtained by integrating over the disc, which extends from the ISCO to infinity, an enhancement of the ISCO radius should lower the luminosity. This can lead to different parameter values for the accreting object and may also involve estimations for the galactic parameters as well. We wish to come back to both of these issues in the future.

As the dark matter halo modifies the geometry, it also has a significant impact on the stability of these galactic wormholes. It turns out that the galactic matter modifies the potential experienced by the perturbing scalar field significantly, by enhancing the throat length but lowering the maxima of the potential. As a consequence, the prompt ringdown will be affected, as the height of the potential decreases, and the echoes in the time domain signal will also be shifted, since the time delay increases. Both of these effects can be seen from \ref{fig-Damour-echo} and \ref{fig-Damour-Log} for the galactic \DS wormhole, while \ref{fig-Brane-echo} and \ref{fig-Brane-Log} depicts such effects for the galactic braneworld wormhole. Interestingly, the imaginary parts of the QNM frequencies for the galactic \DS wormhole are higher compared to the QNM frequencies of the isolated \DS wormhole, indicating that the presence of galactic matter makes the \DS wormhole more stable. On the other hand, for the galactic braneworld wormhole, we have observed the exact opposite scenario. It is expected that, if echoes are observed in the future generations of GW detectors, then a correlation between the parameters derived from the prompt ringdown and from the time delay can provide indirect evidence for the existence of a dark matter halo. This is because, unlike other models of ECOs, for wormholes, the time delay is related to various parameters of the wormhole geometry and hence is not free. Thus, future generations of GW detectors can possibly be used as an independent probe for identifying the galactic parameters.

In that analysis it turned out that the spacetime is linearly unstable due to existence of exponentially growing mode in time. Now coming back to our context. It turns out that the galactic matter modifies the potential experienced by the perturbing scalar field significantly, by enhancing the throat length but lowering the maxima of the potential. As a consequence, the prompt ringdown will be affected, as the height of the potential decreases, and the echoes in the time domain signal will also be shifted, since the time delay increases. Both of these effects can be seen from \ref{fig-Damour-echo} and \ref{fig-Damour-Log} for the galactic \DS wormhole, while \ref{fig-Brane-echo} and \ref{fig-Brane-Log} depicts such effects for the galactic braneworld wormhole. Interestingly, the imaginary parts of the QNM frequencies for the galactic \DS wormhole are higher compared to the QNM frequencies of the isolated \DS wormhole, indicating that the presence of galactic matter makes the \DS wormhole more stable. On the other hand, for the galactic braneworld wormhole, we have observed the exact opposite scenario. It is expected that, if echoes are observed in the future generations of GW detectors, then a correlation between the parameters derived from the prompt ringdown and from the time delay can provide indirect evidence for the existence of a dark matter halo. This is because, unlike other models of ECOs, for wormholes, the time delay is related to various parameters of the wormhole geometry and hence is not free. Thus, future generations of GW detectors can possibly be used as an independent probe for identifying the galactic parameters.

It will be worth to mention it here that the stability analysis reported above is due to a test scalar field, which does not affect the background geometry. It is important to perturb the source of the wormhole itself and check for stability. There have been few works in this direction with mixed results --- (a) wormhole geometries arising out of phantom matter was shown to be unstable under perturbation of the matter sector \cite{Gonzalez:2009hn,Gonzalez:2008wd}, while (b) there are some wormholes in general relativity, which have also been shown to be stable under matter perturbation \cite{Bronnikov:2013coa}. It will be interesting to perform a similar study on the wormhole geometries considered here, which we wish to come back to in a future work.

There are several other future directions of exploration possible. In this work, we have employed the Hernquist-type mass profile for describing the galactic wormhole spacetimes, however, a more complete mass profile, e.g., the NFW profile \cite{Navarro:1995iw}, can also be used. Though, with the use of the NFW mass profile, it would not be possible to obtain any closed-form analytic solutions, they may provide a more realistic description of the impact of dark matter halo on wormhole geometries. Furthermore, we have not discussed the response of these galactic wormhole spacetimes due to external tidal effects. Since it is well-known that ECOs, including wormholes, have non-zero tidal Love numbers, it would be interesting to see how much these love numbers are shifted upon the inclusion of a dark matter halo surrounding the isolated wormholes. Besides, understanding the influence of environmental factors on the physical properties of other classes of ECOs, including quantum-corrected black holes, will be another interesting avenue to explore. Finally, the effect of accretion on the stability of wormhole geometries remains to be understood. In particular, how accretion shifts the quasi-normal mode frequencies and whether it can make the wormhole throat unstable is an interesting direction to explore. We hope to return to these issues elsewhere. 

\section*{Acknowledgment}

SB thanks IACS for financial support and Sayan Kar for the helpful discussion during a meeting. CS thanks the Saha Institute of Nuclear Physics (SINP), Kolkata for financial support. CS is thankful to IACS, for their warm hospitality and research facilities. Research of S.C. is funded by the INSPIRE Faculty fellowship from DST, Government of India (Reg.
No. DST/INSPIRE/04/2018/000893). A part of this work was completed in the Max-Planck Institute for gravitational Physics at Golm, Germany and S.C. acknowledges the warm hospitality there.

\appendix
\labelformat{section}{Appendix #1} 
\labelformat{subsection}{Appendix #1}
\section{An alternative mass profile for Damour-Solodukhin wormhole}\label{app-A}

In the main text, we advocated one particular choice of the mass profile for the galactic \DS wormhole. However, it turns out that there can be other possible choices for the mass profile. In particular, we present below one such alternative mass profile, which reads,
\begin{align}\label{Damour-Gal-Mass-2}
m(r)=M_{2}+\frac{M r^{2}}{(r+a)^{2}}\left(1-\frac{2M_{2}}{r}\right)^{2}~.
\end{align}
Using the above mass profile, one can indeed solve the radial Einstein's equations, as presented by \ref{radial-Einstein}, which yields the following solution for the $g_{tt}$ component of the metric,
\begin{align}\label{Damour-Gal_f}
f(r)=\frac{\exp{G(r)}}{r}~\left(r-2M_{1}\right)^{\frac{(a+2 M_1)^2}{q}} \left(a^2+2 a r+4 M M_2-2 M r+r^2\right)^{\frac{2 M (M_2-M_1)}{q}}~,
\end{align}
where, the function $G(r)$ has the following expression,
\begin{align}
G=\frac{2}{q}\left\{a^2+2 a (M_1+M_2)+2 M (M_2-M_1)+4 M_1 M_2\right\}\sqrt{\frac{M}{\xi }} ~F(r)~;
\quad 
F=\tan^{-1}\left(\frac{a-M+r}{\sqrt{M \xi }}\right)-\frac{\pi }{2}~,
\end{align}
and the constants $\xi$ and $q$ reads as: $\xi=2a-M+4M_{1}$, and $q=(a+2 M_1)^2+4 M (M_2-M_1)$. Note that this function also vanishes at $r=2M_{1}$, but remains non-zero at $r=2M_{2}$, signalling the existence of a wormhole throat at $r=2M_{2}$. Hence the basic nature of the galactic \DS wormhole spacetime remains the same, even with this alternative mass profile. However, due to its complicated functional dependence on the radial coordinate, we have not used it for further analysis. 

\section{The tangential pressure for galactic Damour-Solodukhin wormhole}\label{app-B}

We have described the energy density of the galactic matter supporting the galactic \DS wormhole. However, we have not explicitly mentioned the radial dependence of the tangential pressure associated with the wormhole spacetime. Here, for completeness, we present the tangential pressure of the dark matter halo, which reads,
\begin{align}
p_{\perp}^{(g)}{}_{(ds)} =\frac{1}{16 \pi  r^2} \bigg[&\frac{2 \lambda ^2 r M_{1} }{(r-2 M_{1})^2}+\left(\frac{2 M r}{(r+a)^{2}-2M(r-2M_{1})}+\frac{2 M_{1}}{r-2 M_{1}}\right)
\nonumber
\\
&\times \bigg\{\frac{2 M \left(\lambda ^2 M_{1} (r-a-4 M_{1})+(a+2 M_{1}) (r-2 M_{1})\right)}{(a+r)^3}+\frac{\lambda ^2 M_{1}}{2 \text{M1}-r}\bigg\}\bigg]~.
\end{align}
Note that, in the limit $\lambda \rightarrow 0$, the above expression coincides with \cite{Cardoso:2021wlq}. Moreover, the tangential pressure is positive for any radii $r>2M_{2}$.

\section{Consistency of the gravitational field equations on the brane}\label{app-C}

In this section, we will sketch how one can arrive at the field equation for the radion incarnation $\Phi$, starting from the Bianchi identity. In what follows we will assume that the Planck brane is vacuum, such that the effective Einstein's equations on the visible brane can be written as,
\begin{align}\label{phi-G}
\Phi G_{\mu\nu}=\frac{\kappa^{2}}{\ell}T^{\textrm{vis}}_{\mu\nu}+T^{\Phi}_{\mu\nu}~.
\end{align}
Now operating $\nabla^{\mu}$ on both sides we get,
\begin{align}\label{Contracted-Bianchi}
G_{\mu\nu}\nabla^{\mu}\Phi=\nabla^{\mu}T^{\Phi}_{\mu\nu}~.
\end{align}
To arrive at the previous result we have used the contracted Bianchi identity $\nabla^{\mu}G_{\mu\nu}=0$ and have assumed that $\nabla^{\mu}T^{\textrm{vis}}_{\mu\nu}=0$. Now using the identity,
\begin{align}\label{R-Identity}
[\nabla_{\alpha},\nabla_{\nu}]\nabla_{\mu}\Phi=-R^{\beta}{}_{\mu\alpha\nu}\nabla_{\beta}\Phi~,
\end{align}
one can show that,
\begin{align}\label{C-T-phi}
\nabla^{\mu}T^{\Phi}_{\mu\nu}=R_{\mu\nu}\nabla^{\mu}\Phi-\frac{3\nabla_{\nu}\Phi}{2(1+\Phi)}\gamma~.
\end{align}
Here we have defined the quantity $\gamma$ as,
\begin{align}\label{gamma-def}
\gamma\equiv\nabla^{\alpha}\nabla_{\alpha}\Phi-\frac{1}{2(1+\Phi)}\nabla^{\alpha}\Phi\nabla_{\alpha}\Phi~.
\end{align}
Using \ref{C-T-phi} in \ref{Contracted-Bianchi} we obtain the following relation,
\begin{align}\label{F-1}
\nabla_{\nu}\Phi\left[\frac{3\gamma}{2(1+\Phi)}-\frac{R}{2}\right]=0~.
\end{align}
On the other hand, taking the trace of \ref{phi-G} we obtain,
\begin{align}\label{R-Phi}
-R\Phi=\frac{\kappa^{2}}{\ell}T^{\textrm{vis}}+T^{\Phi}~.
\end{align}
Using the explicit form for the energy-momentum tensor $T^{\Phi}_{\mu\nu}$ of the radion field $\Phi$, one can show that,
\begin{align}\label{trace-Phi}
T^{\Phi}=-3\gamma~.
\end{align}
Therefore, using \ref{trace-Phi} and \ref{R-Phi} in \ref{F-1} we can finally arrive at the following differential equation for $\Phi$,
\begin{align}\label{F-2}
\nabla_{\nu}\Phi\left[-\frac{2\omega+3}{2}\gamma+\frac{\kappa^{2}}{2\ell}T^{\textrm{vis}}\right]=0~,\quad \omega\equiv-\frac{3\Phi}{2(1+\Phi)}~.
\end{align}
The above equation can be satisfied in two possible manners --- (a) if $\Phi$ is a constant field, or, (b) if $\Phi$ satisfies the following differential equation
\begin{align}
\gamma=\frac{\kappa^{2}}{\ell}\frac{T^{\textrm{vis}}}{2\omega+3}~.
\end{align} 
Using the definition of $\gamma$ from \ref{gamma-def}, as well as the coupling function $\omega$, we get back the field equation of the radion incarnation $\Phi$ used in the main text.  

\section{Effective potential for galactic braneworld wormhole}\label{app-D}

The complete expression for the effective potential experienced by a test scalar field living in the galactic braneworld wormhole geometry is given by,
\begin{equation}\label{Brrane-effective-pot}
\begin{split}
V_{l}^{(\textrm{b})} (r) &=\frac{e^{\gamma}}{(1+\chi)^2 r^4 (a+r)^3 \sqrt{1-\frac{2 M_{1}}{r}}}\Bigg((a+r) (r-2 M_{1}) \left(\chi+\sqrt{1-\frac{2 M_{1}}{r}}\right) \times \\ & \left(a^2 M_{1}+2 a M_{1} r+M \left(4 M_{1}^2+\chi r^2 \sqrt{1-\frac{2 M_{1}}{r}}-4 M_{1} r+r^2\right)+M_{1} r^2\right) \\& +  r \sqrt{1-\frac{2 M_{1}}{r}} \left(\chi+\sqrt{1-\frac{2 M_{1}}{r}}\right)^2  \Bigg(a^3 M_{1}+3 a^2 M_{1} r+a \left(4 M M_{1}^2-M r^2+3 M_{1} r^2\right) \\& +M r \left(12 M_{1}^2-8 M_{1} r+r^2\right)+M_{1} r^3\Bigg) +l (l+1) r^2 (a+r)^3 \sqrt{1-\frac{2 M_{1}}{r}} \left(\chi+\sqrt{1-\frac{2 M_{1}}{r}}\right)^2\Bigg)~.
\end{split}
\end{equation}
As in the case of galactic \DS wormhole spacetime, in this case as well the effective potential $V_{l}^{(\textrm{b})} (r)$ can be approximated as follows,
\begin{align}\label{Brane-pot-app}
V_{l}^{(\textrm{b})} (r)\simeq \frac{e^{\gamma}}{r^{4}(a+r)^{3}} \left[V_{0}+\chi V_{\chi}\right]~,
\end{align}
where $V_{0}$ is the leading order term independent of the wormhole parameter $\chi$ and $V_{\chi}$ is the coefficient of the linear order term in $\chi$, which are defined as follows,
\begin{align}\label{brane-vo}
 & V_{0}= (a+r) (r-2 M_{1}) \bigg(a^2 M_{1}+2 a M_{1} r+2 M M_{1} (2 M_{1}-r)+M r (r-2 M_{1})+M_{1} r^2\bigg)+\nonumber\\
& r \bigg(1-\frac{2 M_{1}}{r}\bigg)\bigg(a^3 M_{1}+3 a^2 M_{1} r+a (4 M M_{1}^2-M r^2+3 M_{1} r^2)+M r (12 M_{1}^2-8 M_{1} r+r^2)+M_{1} r^3\bigg)+\nonumber\\
&l (l+1) r^2 (a+r)^3 \bigg(1-\frac{2 M_{1}}{r}\bigg)~,
\end{align}
and,
\begin{align}\label{brane-chi}
 &V_{\chi}=(a+r) (r-2 M_{1}) \bigg(a^2 M_{1}+2 a M_{1} r+2 M M_{1} (2 M_{1}-r)+2 M r (r-2 M_{1})+M_{1} r^2\bigg)-\nonumber\\
 &2 \sqrt{1-\frac{2 M_{1}}{r}} (r-2M_{1}) \bigg(a^3 M_{1}+(a+r) (a^2 M_{1}+2 a M_{1} r+2 M M_{1} (2 M_{1}-r)+M r (r-2 M_{1})+M_{1} r^2)+\nonumber\\
 &3 a^2 M_{1} r+l (l+1) r (a+r)^3+a (4 M M_{1}^2-M r^2+3 M_{1} r^2)+M r (12 M_{1}^2-8 M_{1} r+r^2)+M_{1} r^3\bigg)+\nonumber\\
 &2 (r-2 M_{1}) \bigg(a^3 M_{1}+3 a^2 M_{1} r+a (4 M M_{1}^2-M r^2+3 M_{1} r^2)+M r (12 M_{1}^2-8 M_{1} r+r^2)+ M_{1} r^3\bigg)+\nonumber\\
 &2 l (l+1) r (a+r)^3 (r-2 M_{1})~. 
\end{align}
These expressions involving the effective potential experienced by a scalar field have been used in \ref{fig-Brane-Pot}, in the main text. 

\bibliography{WormGal}

\bibliographystyle{utphys1}
\end{document}